\documentclass[sigconf,nonacm]{acmart}
\usepackage[most]{tcolorbox} 
\usepackage{multirow}
\usepackage{svg}
\usepackage[skip=1pt]{caption}

\usepackage[table,xcdraw]{xcolor}
\newcommand{\approach}{\textsc{CALM }}

\tcbset{
  examplebox/.style={
    colback=blue!15!white,
    colframe=blue!40 !white,
    boxrule=0.5pt,
    arc=2mm,
    left=1mm,
    right=1mm,
    top=1mm,
    bottom=1mm,
    fonttitle=\bfseries,
  },
  summarybox/.style={
    colback=cyan!18!white,
    colframe=cyan!200 !white,
    boxrule=0.5pt,
    arc=2mm,
    left=1mm,
    right=1mm,
    top=1mm,
    bottom=1mm,
    fonttitle=\bfseries,
    fontupper=\small
  }
}

\AtBeginDocument{%
  }

\acmConference[SEAMS '26]{21st International Conference on Software Engineering for Adaptive and Self-Managing Systems}{April 13--14, 2026}{Rio de Janeiro, Brazil}
\acmBooktitle{21st International Conference on Software Engineering for Adaptive and Self-Managing Systems (SEAMS '26), April 13--14, 2026, Rio de Janeiro, Brazil}
\acmDOI{10.1145/3788550.3794880}
\begin{document}


\title{CALM: \textcolor{black}{A Self-Adaptive Orchestration Approach for QoS-Aware Routing in Small Language Model based Systems}}

\author{Hemang Jain}
\authornote{Both authors contributed equally.}
\email{hemang.jain@students.iiit.ac.in}
\affiliation{%
  \institution{IIIT Hyderabad, India}
  \country{India}
}

\author{Divyansh Pandey}
\authornotemark[1]
\email{divyansh.pandey@students.iiit.ac.in}
\affiliation{%
  \institution{IIIT Hyderabad, India}
  \country{India}
}

\author{Karthik Vaidhyanathan}
\email{karthik.vaidhyanathan@iiit.ac.in}
\affiliation{%
  \institution{IIIT Hyderabad, India}
  \country{India}
}

\begin{abstract}
\textcolor{black}{AI-enabled systems are
subjected to various types of runtime uncertainties}, ranging from dynamic workloads, resource requirements, model drift, etc. These uncertainties have a big impact on the overall Quality of Service (QoS). This is particularly true in the case of Language Model (LM) enabled systems where the autoregressive nature of token generation introduces variability in latency, energy usage and response quality. These systems, powered by LLMs, are either resource-intensive (if run on-prem) or raise privacy/cost concerns (if leveraged using APIs). While deploying a Small Language Model (SLM) can be resource-efficient, it often falls short in addressing the diversity and scale of real-world requirements. To this, we argue that, rather than relying on any one SLM, leveraging a coordinated fleet of SLMs, each with specialized strengths can enable systems to dynamically adapt to shifting contexts and workload patterns. However, realizing the full potential of such an approach demands intelligent orchestration and continuous adaptation. To this end, we introduce \approach, a self-adaptive orchestration mechanism based on MAPE-K. Our approach continuously monitors user queries, \textcolor{black}{analyzes the QoS metrics of the SLMs}, \textcolor{black}{identifies the optimal SLM to be used, 
routes the query to the identified SLM and further to enhance the effectiveness and efficiency, leverages caching and scheduling} to decide the SLMs to be kept in memory. Our evaluation shows that CALM reduces latency by approximately \textbf{40\%} and energy consumption by \textbf{50\%}, while preserving domain-specific task performance when compared to single-LLM baselines. 
\footnote{Github Link: https://github.com/sa4s-serc/CALM}
\end{abstract}

\maketitle
\section{Introduction}
\label{sec:intro}

Modern software systems operate in dynamic and uncertain environments. These uncertainties can arise not only from external factors such as workload fluctuations or resource availability but also from within the software itself. This is especially true for AI-enabled systems. As noted by Lewis et al.~\cite{lewis2024software}, uncertainty is an inherent property of AI systems, emerging from data, models, or contextual variability. Over the years, several works have explored self-adaptation as a mechanism to handle such uncertainties, enabling systems to monitor their behavior and autonomously adjust to maintain desired Quality of Service (QoS).

The rapid growth of generative AI has further magnified these challenges. Large Language Models (LLMs) such as GPT-4, LLaMA, Gemini, and DeepSeek have shown remarkable capabilities in generating coherent and contextually rich text~\cite{openai2024gpt4technicalreport,grattafiori2024llama3herdmodels,geminiteam2025geminifamilyhighlycapable,deepseekai2025deepseekv3technicalreport}. However, their auto-regressive nature, where each generated token depends on training corpus and input context, introduces additional non-determinism. The latency, accuracy, energy and memory footprint of these models vary significantly depending on architecture, size, and number of output tokens. Moreover, factors such as prompt clarity, query arrival rate, and deployment setup further add to runtime uncertainties, impacting  model-level and system-level QoS.

While large general-purpose LLMs offer broad capabilities ~\cite{openai2024gpt4technicalreport,grattafiori2024llama3herdmodels,geminiteam2025geminifamilyhighlycapable,deepseekai2025deepseekv3technicalreport}, they are often over-engineered for domain-specific applications like finance, healthcare, or law~\cite{10.1145/3697012,jayakumar2023largelanguagemodelslegal,wu2023bloomberggptlargelanguagemodel}. Recent studies have demonstrated that when operating under limited capabilities, LLMs tend to favour fluent and persuasive responses over accurate, domain-specific expertise, potentially misleading users with surface-level competence \cite{wysocka2024factuality}. Moreover, they incur high computational costs and API charges~\cite{irugalbandara2024slam,balloccu2024leakcheatrepeatdata}. In contrast, smaller, domain-specialised Small Language Models (SLMs) can provide lower latency, reduced computational cost, and higher task accuracy ~\cite{subramanian2025smalllanguagemodelsslms} for well-defined contexts. However, in such settings, it becomes important that the system can dynamically decide which SLM to invoke for a given query, especially when multiple SLMs are available with different capabilities and resource needs. 
This can be enabled via a self-adaptive mechanism that continuously observes system conditions and autonomously selects the most suitable model at runtime, allowing system to balance latency, accuracy, and resource usage as conditions evolve.
However, enabling such adaptive orchestration brings several software engineering challenges: identifying which model to use for a given request, keeping frequently used models loaded in memory to reduce inference delays, managing GPU constraints through efficient caching, and scheduling inference queries to meet QoS objectives. Although some approaches have explored model switching at runtime, they often overlook the overheads and constraints specific to auto-regressive language models~\cite{casimiro2021self,kulkarni2023towards}.

To address these challenges, we present CALM, a self-adaptive orchestration approach for efficient inference across multiple specialized language models. CALM integrates routing, scheduling, and caching within a MAPE-K~\cite{7194653} feedback loop: the Monitor collects runtime data, the Analyzer evaluates QoS adherence, the Planner updates routing and scheduling strategies, and the Executor enacts them, while Knowledge maintains system goals and policies. Through this adaptive loop, CALM dynamically selects the most suitable model, optimizes resource utilization, and continuously adapts to varying workloads and contexts. Our evaluation across diverse tasks shows that CALM outperforms single-LLM baselines in latency, efficiency, and confidence demonstrating how self-adaptation can effectively enable stable and cost-efficient operation of modern AI systems.

\begin{figure*}[htbp]
\caption{ \approach with Self-Adaptive Loop \& Cache Module for Routing \& Scheduling queries across Fleet of SLMs.}
\centerline{\includegraphics[width=0.88\linewidth]{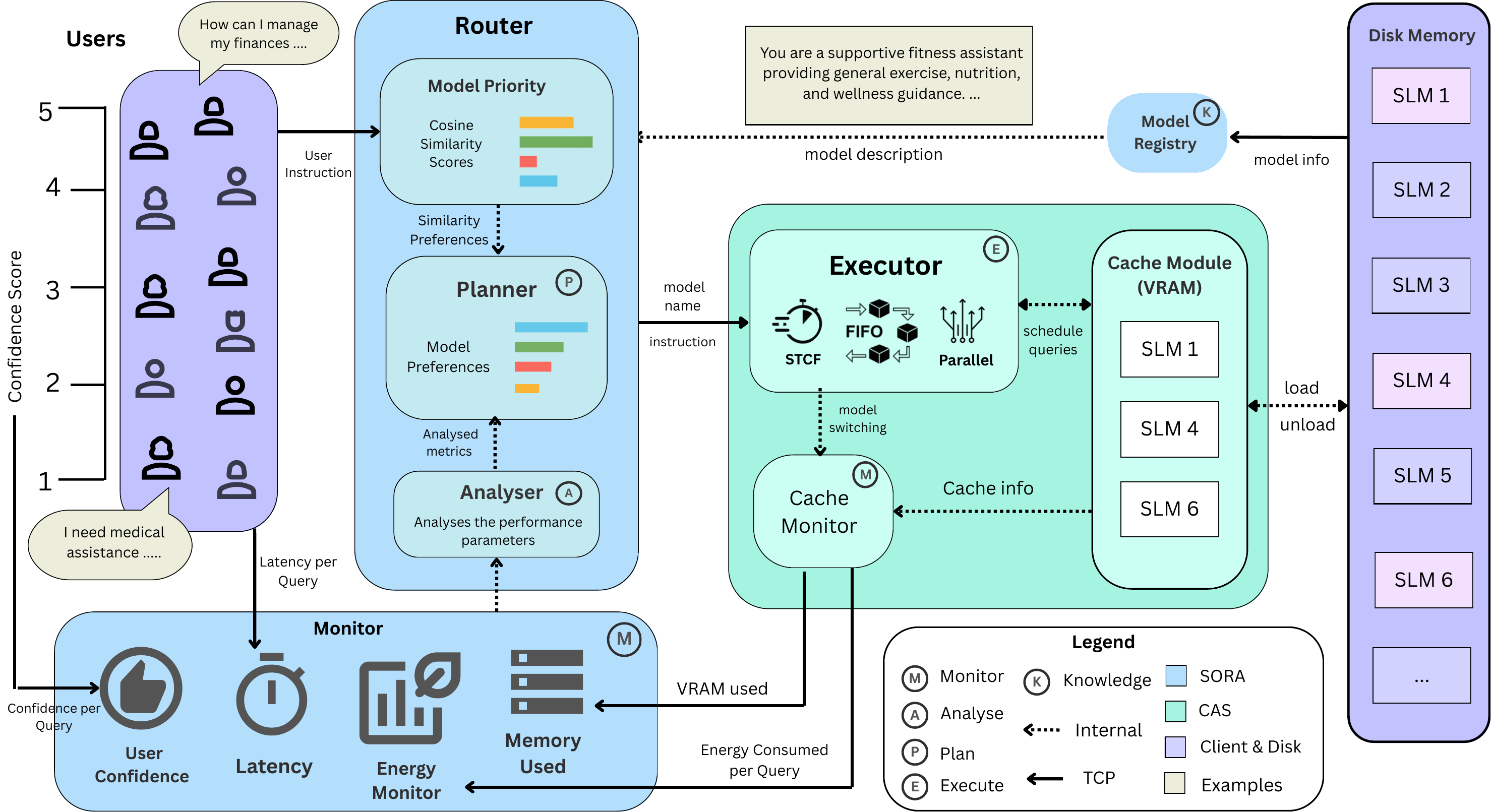}}
\label{fig:main}
\end{figure*}


\vspace{-7pt}
\section{Problem Statement}
\color{black}
While LLMs demonstrate strong general-purpose language understanding, they present notable challenges in real-world deployments. Their high computational cost and limited domain-specific factual accuracy call for the need of specialized, domain-specific SLMs. The following challenges remain critical.
\subsubsection*{Challenge 1:}
   LLMs often fall short in domain-specific tasks, as they tend to favor generic, appealing responses over factually accurate ones ~\cite{choi-etal-2024-llm}. To achieve high factual accuracy in critical domains, LLMs require fine-tuning on domain data, which is resource-intensive due to their massive size. SLMs, by contrast, require significantly fewer resources for tuning and can achieve comparable accuracy in domain-specific tasks.
\subsubsection*{Challenge 2:}
   Even after such domain adaptation, running LLMs, whether hosted on premise or via APIs, requires substantial memory and compute resources along with major environmental impact.
\color{black} \subsubsection*{Challenge 3:} \color{black} Current self-adaptive routing approaches like SWITCH \cite{marda2024switch} work well with ML and vision tasks but do not adapt to the uncertainties associated with Language Models. In LMs, the auto-regressive nature of text generation causes the number of output tokens to vary significantly with model architecture, fine-tuning, and input semantics. This uncertainty impacts not only the confidence scores, but also the latency and energy consumption per query. Furthermore, naïve load-balancing strategies often fail to adapt effectively to the inherent semantic dependencies.\\ \\ 
    \color{black} To overcome the above challenges, replacing a single LLM with a fleet of domain-specialized SLMs can be a potential candidate, however, 
    deploying SLMs at scale presents some key challenges. Keeping all models in memory simultaneously is impractical due to hardware limitations. Additionally, incoming queries must be routed intelligently to the most appropriate model, ensuring high QoS while respecting resource limits.

\color{black}
 To address these challenges we develop \textbf{\approach}, guided by the hypothesis that a fleet of domain-specialized SLMs, when orchestrated intelligently, can outperform single, general purpose LLMs on domain-specific tasks with improved efficiency and controllability. Formally, given a set of user queries denoted by $\mathcal{Q}$, a set of domains $\mathcal{D}$, and a corresponding collection of models $\mathcal{M} = \{m_i\}_{i=1}^N$, where each model $m_i$ is specialized for a domain $d_i \in \mathcal{D}$, our goal is to \textit{route each query} $q \in \mathcal{Q}$ to the most appropriate model $m^* \in \mathcal{M}$ in a way that maximizes overall QoS, balancing \textbf{accuracy, latency, and resource efficiency} under hardware and memory constraints. To optimize query routing, we introduce a self-adaptive MAPE-K loop that monitors model performance over time and maintains rolling metrics. We combine semantic similarity with rolling performance metrics to prioritize models, and route each query $q \in \mathcal{Q}$ to the model $m^*$ that offers the best trade-off across both latency and confidence. Additionally, to reduce memory usage, we design a cache policy $\pi_c$ based on model access patterns and system constraints.

\vspace{-7pt}
\color{black}
\section{The CALM Approach}
\color{black}

Our approach is designed to enable efficient, adaptive, and resource-aware serving of multiple domain-specialised SLMs. To meet the demands of low latency, energy efficiency, and high accuracy under constrained resources, our design comprises two key components: SORA and CAS  integrated within a self-adaptive feedback loop.

    \textbf{SORA (Semantic-aware Orchestrator}: Maintains a registry of available SLMs, performs semantic-aware query routing, and coordinates execution.
    
    \textbf{CAS (Cache-enabled Adaptive Scheduler)}: Handles dynamic loading and eviction of models through a cache module that retains a fixed number of models in GPU memory \textcolor{black}{(referred as VRAM)}. It is also responsible for scheduling queries and executing inference.
    
     \textbf{Self-Adaptive Feedback Loop}: Implements a MAPE-K loop to adapt routing decisions based on real-time performance metrics.

As illustrated in Figure~\ref{fig:main}, SORA and CAS work together to process user queries end-to-end. Sequentially, before serving users, SORA registers available SLMs. After receiving a user request, the router ranks the models based on semantic similarity between the query and model descriptions, as well as on the run-time metrics collected by the Monitor. Based on a routing policy, the appropriate model is selected, and the query is forwarded to the executor. If the selected model resides in \textcolor{black}{VRAM}, it is used directly; otherwise, it is loaded into the Cache Module on demand. The caching policy is motivated to be designed to reduce the frequency of cold starts and self-adaptive routing provides more control on their occurrence \textcolor{black}{(as we demonstrate in our implementation later in Section 5)}. The response is returned to the user, who provides a confidence score, and the Monitor logs the observed latency of the end-to-end pipeline per query. CAS simultaneously records energy consumption and memory usage during inference to support continuous adaptation.

Throughout the section, we use a running example query from the health care domainconfined to illustrate our approach. A fleet of specialized SLMs is pre-registered with the SORA, these include, for instance, a flu diagnosis model, an Ear, Nose, and Throat (ENT) specialist model, a Covid-19 specialist model, and a general medical assistant model. CALM, comprising the Router, Scheduler and the fleet of SLMs, is exposed to users as a unified interface. Each subsequent subsection demonstrates how the example query is handled at each stage of the approach, to ground the technical details in a real-world scenario.



\subsection{Model Registration and Representation}

While registering, each model \( m_i \in \mathcal{M} \), where \( \mathcal{M} \) denotes the set of all registered models, submits a metadata tuple \( (d_i, \theta_i) \) to the SORA registry. Here:

\begin{itemize}
    \item \( d_i \) is the model's semantic task description — a natural language specification (e.g., system prompt) that captures the model’s fine-tuned purpose or domain of expertise. This information is essential for identifying models relevant to a given user query.

    \item \( \theta_i \) comprises model-specific attributes required for communication and execution, including but not limited to the model's identifier, name, IP address, port, and other operational metadata.
\end{itemize}

The SORA maintains a dynamic registry of these tuples as shown in the Model registry in Figure ~\ref{fig:main}, to support effective routing and scheduling. The semantic component \( d_i \) enables query-to-model relevance matching, while \( \theta_i \) ensures the SORA has sufficient physical detail to dispatch requests to the appropriate model instance. The next step involves determining the appropriate model to route this query to as discussed in next section.

\begin{tcolorbox}[examplebox]
\textbf{Example}: A flu diagnosis model, an SLM fine-tuned on flu-related queries and responses, registers with the description “You are a flu assistant that provides accurate diagnoses based on symptom descriptions,” along with IP \texttt{192.168.1.10} and port \texttt{9000}. Other registered models include a general medical diagnosis model, ENT specialist, pediatric symptom checker, and COVID-19 screener. A user submits a query describing throat pain and mild fever and seeks guidance. 
\end{tcolorbox}

\subsection{Self-Adaptive Query Routing}

{\color{black}SORA employs a self-adaptive routing policy $\pi_r$ to select the most suitable model 
$m^* \in \mathcal{M}$ for a given query $q$. The routing policy evaluates each candidate model 
using a combination of static semantic relevance and dynamic runtime performance, as governed 
by system objectives encoded in the MAPE-K loop.

Formally, for a query $q$ and model $m_i$, the routing policy assigns a routing score:
\[
\pi_r(q, m_i) = \texttt{score}(q, m_i).
\]

\paragraph{Static Score.}
The static score $\texttt{similarity\_score}(q, d_i)$ captures the semantic alignment between 
the query $q$ and the task description $d_i$ of model $m_i$, using an appropriate similarity 
metric (e.g., cosine similarity~\cite{reimers2019sentencebertsentenceembeddingsusing}). 
Depending on the routing objective, this score may be used directly (to be maximized) or 
inverted as $(1 - \texttt{similarity\_score})$ (to be minimized).

\paragraph{Dynamic Score.}
The dynamic score reflects rolling runtime performance metrics collected by the 
\textbf{Monitor} phase of the MAPE-K loop during inference. SORA monitors one or more runtime 
metrics, such as latency, energy consumption, peak VRAM usage, or model confidence. For each 
metric $l$ and model $m_i$, a metric-specific dynamic score is computed as:
\[
\texttt{dynamic\_score}_l(m_i) =
\frac{1}{k} \sum_{j=1}^{k} \widehat{\texttt{metric}}_l^{(j)}(m_i),
\]
where $k$ is the number of recent observations and 
$\widehat{\texttt{metric}}_l(\cdot)$ denotes a normalized metric value scaled to $[0,1]$.

\paragraph{Routing Score Computation.}
The \textbf{Analyze} phase interprets recent performance trends, while the \textbf{Plan} phase 
selects a set of tunable control parameters 
$\{\lambda_l\}$ based on high-level objectives stored in the \textbf{Knowledge} component 
(e.g., latency-sensitive, energy-efficient, or confidence-driven routing). These parameters 
determine how strongly each dynamic metric influences routing decisions.

The final routing score is computed as:
\begin{equation}
\begin{aligned}
\texttt{score}(q, m_i) =\;&
g\!\left(\texttt{similarity\_score}(q, d_i)\right) \\
&+ \sum_{l} \lambda_l \cdot \texttt{dynamic\_score}_l(m_i)
\end{aligned}
\end{equation}

where $g(\cdot)$ is either the identity function or an inversion 
$(1 - \cdot)$, depending on whether the routing objective is formulated as a maximization or 
minimization problem

\paragraph{Handling Correlated metrics.}
Runtime metrics may exhibit correlation (e.g., latency and energy consumption), which can bias
linear aggregation by overemphasizing redundant signals.
While the current implementation assumes independent metric contributions, the routing
formulation naturally supports correlation-aware weighting.
Let $\rho(l,l')$ denote the empirical correlation between (say Spearman correlation) metrics $l$ and $l'$ computed over
recent observations.
The influence of metric $l$ can be attenuated as:
\[
\lambda_l' =
\frac{\lambda_l}{1 + \sum_{l' \neq l} |\rho(l,l')|}.
\]
The routing score can then be computed using the adjusted weights:
\begin{equation}
\begin{aligned}
\texttt{score}(q, m_i) =\;&
g\!\left(\texttt{similarity\_score}(q, d_i)\right) \\
&+ \sum_{l \in \mathcal{L}} \lambda_l' \cdot
\texttt{dynamic\_score}_l(m_i).
\end{aligned}
\end{equation}

This reduces the impact of correlated metrics without altering the core routing
mechanism. For the scope of this work, we do not aim to handle correlated metrics or multiple metrics optimization, an empirical evaluation of which is left for future work.

\paragraph{Model Ranking and Selection.}
Models are ordered using a ranking operator $\texttt{rank}(\cdot)$, which sorts candidate 
models according to their routing scores in ascending or descending order, consistent with 
the chosen objective. The selected model is:
\[
m^* = \arg\min_{m_i \in \mathcal{M}} \texttt{rank}\!\left(\pi_r(q, m_i)\right).
\]

After the router selects the most suitable model for a query, the request is prepared for 
execution. The query is bundled with the selected model information and handed to the 
executor (Fig.~\ref{fig:main}), which manages request scheduling as described below.
}

\begin{tcolorbox}[examplebox]
 \textbf{Example}: 
A user query like “I have a sore throat and mild fever” is routed to the flu model, which returns a diagnostic suggestion (e.g., rest, hydration). If the flu model’s latency increases over time due to GPU congestion or hardware degradation the MAPE-K loop detects the issue via continuous monitoring. Subsequent similar queries may then be routed to alternative models, such as a COVID-19 model, selected based on semantic similarity and performance metrics to ensure both relevance and efficiency.

\end{tcolorbox}

\subsection{Query Processing and Scheduling}

Given a user query \( q \), a request tuple \( r_j = (q_j, m_i, u_j) \) is provided to the executor, where:
\begin{itemize}
    \item \( q_j \): the user query content,
    \item \( m_i \in \mathcal{M} \): the model selected by the router as the most relevant to \( q_j \),
    \item \( u_j \): user-specific metadata required for response dispatch (e.g., client ID, callback address).
\end{itemize} 

The executor in CAS maintains a global request queue \( \mathcal{Q} = [r_1, r_2, \ldots, r_n] \), where each element \( r_j \) represents a routed request. This queue serves as a centralized buffer that decouples query arrival from execution, 
\textcolor{black}{enabling the system to admit and manage requests from multiple users and models simultaneously.}

A dedicated scheduler operates over \( \mathcal{Q} \), applying a configurable policy to select the next request to serve (representing the \textbf{Execute} in the MAPE-K loop). Common scheduling policies include:

     \textbf{First-In-First-Out (FIFO):} Requests are processed in order of arrival. The scheduler selects the earliest unprocessed \( r_j \in \mathcal{Q} \).
    
     \textbf{Shortest-Time-to-Completion-First (STCF):} Prioritizes requests based on the expected inference time of the assigned SLM calculated as average of inference times of past served requests. Formally, the next request is:
    \[
    r_j = \arg\min_{r_k \in \mathcal{Q}} \text{inftime}(m_k),
    \]
    where \( m_k \) is the SLM prioritized by router for the request \( r_k \), and \( \text{inftime}(m_k) \) denotes SLM's expected inference time.

\begin{tcolorbox}[examplebox]
\textbf{Example}: SORA packages the query “sore throat and fever” into a request tuple with the selected flu model and user metadata, then appends it to the global queue. The scheduler processes requests based on the configured policy, for example, FIFO ensures fairness, while STCF prioritizes quick completions, favoring lightweight models.
 
\end{tcolorbox}

\subsection{Caching Mechanism}
To provide users with the illusion of seamless access to a large pool of models, the CAS employs a memory-aware Cache Module that dynamically maintains only a subset of models in main memory at any given time. Let \( C \subseteq \mathcal{M} \) represent the current cache set, with \( |C| \leq k \), where \( k \) is the cache capacity determined by available memory. Models outside this set reside in secondary storage and must be loaded before inference. This management remains transparent to users, who perceive all models as instantly accessible.

When a requested model \( m^* \notin C \), the system must either place it into free space or evict an existing model \( m_e \in C \) according to an \textit{eviction policy}, denoted by \( \pi_c \). The choice of policy \( \pi_c \), along with the cache size \( k \), directly impacts the cache hit rate, influencing how often expensive model load operations occur. A higher hit rate leads to lower latency and more responsive performance.

Eviction policies prioritize which models remain in memory. A Least Recently Used (LRU) policy, for instance, favors models accessed in the recent past and adapts well to shifting user interests. In contrast, a Least Frequently Used (LFU) policy retains models with high long-term access frequency, suiting workloads with stable usage patterns. Such policies enable efficient memory use even in constrained environments, supporting multiple models without exceeding memory budgets.

\begin{tcolorbox}[examplebox]
\textbf{Example}: If the flu model isn’t cached at scheduling time, the system checks for free memory. If full, it uses the eviction policy—e.g., LRU may evict the least-used model. The flu model is then loaded from storage into memory. 

\end{tcolorbox}

\subsection{Execution and Response}

Once a query \( q \) is routed to the selected model \( m^* \), the system ensures that \( m^* \) is loaded into memory, either directly from the cache \( C \) or by fetching it from secondary storage if necessary. After ensuring model availability, the CAS layer executes the model's inference, generating a response \( r = m^*(q) \).

The generated result is then relayed back to the SORA, which uses metadata from the query (such as user context and request tracking information) to route the response to the correct client. The client has an option to rate the response generated from 1-5. This feedback is recorded for maintaining average confidence values of the models hosted by CAS to track their performance.

\begin{tcolorbox}[examplebox]
\textbf{Example}: The loaded flu model processes the query and returns a diagnostic response suggesting rest and hydration. SORA forwards this to the user via the the IP Address and Port. The user rates it 4 out of 5, and this feedback updates the flu model’s average confidence score for future routing and quality tracking.

\end{tcolorbox}

\section{Experimentation Design}

%
%

Our experimental design is established with the aim to answer the following research questions:\\\\
\textbf{\underline{RQ1:} \space How effective is the proposed approach in optimizing task performance?}  \underline{Aim}: To assess the effectiveness of system to deliver better QoS with efficient resource usage. For a comprehensive analysis, we break it down into the following sub-questions:
    \begin{itemize}
        \item[\underline{RQ1.1:}] How does the system perform across key metrics—latency, confidence score, energy consumption, when compared to baseline LLMs?
        \item[\underline{RQ1.2:}] How does the cache module impact memory utilization while maintaining QoS?
        \item[\underline{RQ1.3:}] How can a self-adaptive MAPE-K loop improve Quality of Service (QoS) by dynamically selecting models by incorporating real-time performance metrics?
    \end{itemize} 
\textbf{\underline{RQ2:} \space How well does the system scale to different workloads and generalize to diverse domains?} \underline{Aim}: To evaluate the system’s ability to handle increasing user load without compromising performance and to demonstrate robustness in multi-domain, real-world settings with varying compositions of SLMs.\vspace{5pt}\\
\textbf{\underline{RQ3:} \space Is the system efficient in terms of decision overhead and runtime feasibility?} \underline{Aim}: To assess whether the system can make adaptive routing and caching decisions with minimal overhead, ensuring low-latency inference.



\subsection{Configuration Selection}

\subsubsection{Workload Simulation}
\begin{figure}[htbp]
\vspace{-10pt}

\caption{Workload Distribution for a 20 minute window}
\centerline{\includegraphics[width=0.9\columnwidth]{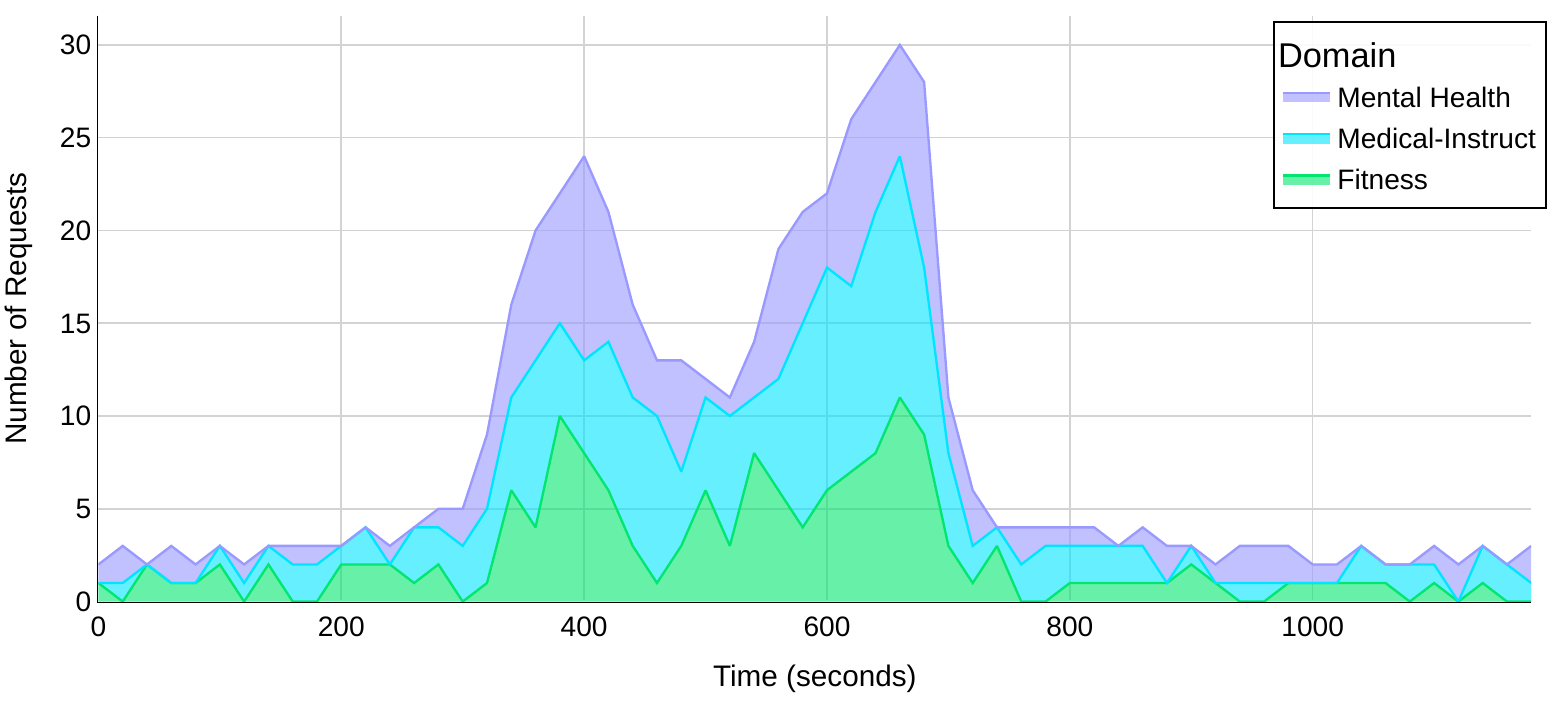}}
\label{fig:workload}
\vspace{-10pt}
\vspace{1mm}
\captionsetup{justification=justified, skip = 3 pt}
\caption*{\footnotesize Caption:\textnormal{Figure shows distribution of 500 queries evenly split across 3 domains. Same distribution is used for inter-domain setting and scaled to 40 minutes for 40 | 256.}}
\end{figure}

To evaluate the performance across different candidates, we consider two maximum output token settings of 256 and 512 tokens to capture the effect of response length on energy consumption, latency and user confidence. In an attempt to simulate realistic user interaction patterns, we simulate 500 user requests, evenly distributed across subdomains and temporally sampled based on the access distribution of the 1998 FIFA World Cup \cite{fifaworldcup}. For the content of these simulated user queries, we draw examples from instruction-based datasets curated for each domain (See 4.1.2). These requests are evaluated under two load conditions: \textbf{12.5} and \textbf{25} requests per minute (RPM) on average, corresponding to 40-minute and 20-minute intervals, respectively. For brevity, we denote workloads as \texttt{T\,|\,L}, where \texttt{T} is the duration in minutes (e.g., 20 or 40) and \texttt{L} is the output token limit (e.g., 256 or 512). For example, \texttt{20\,|\,256} refers to a 20-minute user-activity window with 256 tokens as maximum response length.(Refer Figure~\ref{fig:workload}). Unless explicitly mentioned, we work with the \texttt{20\,|\,256} workload.
\subsubsection{Domain Selection}
To evaluate the effectiveness and generalization ability of \approach  approach
 we consider two distinct domain selection settings. The first comprises of models fine-tuned on related subdomains within healthcare: \textbf{physical fitness}, \textbf{clinical medicine}, and \textbf{mental health}. These subdomains vary in task structure, ranging from clinical fact-based QA to empathetic conversational support—allowing us to assess specialization within semantically proximate areas.
The second setting covers disjoint domains selected for their diversity: \textbf{law}, \textbf{financial services}, and \textbf{clinical healthcare}.
This setting enables us to study model behavior across heterogeneous and non-overlapping domains.

We selected instruction-based datasets on the same to fine-tune the models. More details of the dataset can be found at \href{https://anonymous.4open.science/r/CALM-1FD3}{github}.

\vspace{-6pt}
\subsubsection{Model Selection and Fine-Tuning}

For each domain, we select multiple instruct SLMs. Since there is no strict definition of SLMs, we adopt a practical threshold and define SLMs as general-purpose models in the 1B--4B parameter range. This reflects a balance between deployability on resource-constrained hardware and sufficient capability for task-oriented reasoning, consistent with how such models are increasingly used in real-world applications. These models are chosen to ensure diversity across architectures, public availability for reproducibility, and instruction tuning to align with task-oriented queries. Specifically, we use instruct-tuned models \textbf{Qwen2.5-3B}, \textbf{Phi-3-mini-4k}, \textbf{LLaMA-3.2-3B}, \textbf{LLaMA-3.2-1B}, and \textbf{Gemma-2-2B}, all evaluated with \texttt{fp16} precision. Each model is assigned a system description, used for computing query similarity scores.
\color{black}All domain-specific SLMs are Supervised Fine-Tuned (SFT) with parameter-efficient LoRA adapters applied to their respective base LLMs. We use the HuggingFace TRL's SFTTrainer with mixed-precision training and gradient checkpointing enabled to reduce VRAM overhead. Training is performed on curated domain-specific instruction–response datasets using a consistent prompt template, a per-device batch size of 2 with gradient accumulation over 4 steps, a learning rate of $1*10^{-3}$, linear learning-rate scheduling with a 3\% warmup ratio, and 2 training epochs. LoRA adapters are applied to attention, feed-forward layers while keeping base model frozen.
\color{black}
\begin{table}[]

\renewcommand{\arraystretch}{1.5}
\caption{List of Models and their Domain Specialization}
\resizebox{\columnwidth}{!}{
\begin{tabular}{|ll|ll|ll|}
\hline
\multicolumn{2}{|c|}{\textbf{Set 1}}                                                 & \multicolumn{2}{c|}{\textbf{Set 2}}                                                 & \multicolumn{2}{c|}{\textbf{Set 3}}                                                 \\ \hline
\multicolumn{1}{|c|}{Model Name}                       & \multicolumn{1}{c|}{Domain} & \multicolumn{1}{c|}{Model Name}                       & \multicolumn{1}{c|}{Domain} & \multicolumn{1}{c|}{Model Name}                       & \multicolumn{1}{c|}{Domain} \\ \hline
\multicolumn{1}{|l|}{gemma-2-2b-it}             & Medical                     & \multicolumn{1}{l|}{gemma-2-2b-it}             & Medical                     & \multicolumn{1}{l|}{gemma-2-2b-it}             & Medical                     \\
\multicolumn{1}{|l|}{gemma-2-2b-it}             & Fitness                     & \multicolumn{1}{l|}{Qwen2.5-3B-Instruct}              & Fitness                     & \multicolumn{1}{l|}{gemma-2-2b-it}             & Legal                       \\
\multicolumn{1}{|l|}{gemma-2-2b-it}             & Mental Health               & \multicolumn{1}{l|}{Llama-3.2-1B-Instruct} & Mental Health               & \multicolumn{1}{l|}{Llama-3.2-1B-Instruct} & Finance                     \\
\multicolumn{1}{|l|}{Llama-3.2-3B-Instruct} & Medical                     & \multicolumn{1}{l|}{Phi-3-mini-4k-instruct} & Medical                     & \multicolumn{1}{l|}{Phi-3-mini-4k-instruct} & Medical                     \\
\multicolumn{1}{|l|}{Llama-3.2-3B-Instruct} & Fitness                     & \multicolumn{1}{l|}{Llama-3.2-1B-Instruct} & Fitness                     & \multicolumn{1}{l|}{Phi-3-mini-4k-instruct} & Finance                     \\
\multicolumn{1}{|l|}{Llama-3.2-3B-Instruct} & Mental Health               & \multicolumn{1}{l|}{Llama-3.2-3B-Instruct} & Mental Health               & \multicolumn{1}{l|}{Phi-3-mini-4k-instruct} & Legal                       \\ \hline
\end{tabular}
}
\label{tab:sets}
\vspace{-5 pt}
\end{table}
\vspace{-3pt}
\subsubsection{Experimental Set Configurations}

To systematically evaluate the system under varied domain and model conditions, we define three experimental sets:

\textbf{Set 1} assesses the impact of domain specialization by fine-tuning a fixed set of SLMs on related subdomains: \textbf{physical fitness, clinical medicine} and \textbf{mental health}. These domains are semantically similar but differ in task structure.

\textbf{Set 2} explores model-agnostic behavior by replacing the models from Set 1 with a new selection from different model families, while keeping the domain set fixed. This helps to evaluate whether our routing and performance trends generalize across different SLMs.

\textbf{Set 3} investigates domain diversity by fine-tuning SLMs on a disjoint set of domains: \textbf{law}, \textbf{financial services}, and \textbf{clinical healthcare}. These domains differ significantly in language style, factual grounding, and task framing, enabling us to evaluate system performance across heterogeneous and non-overlapping areas.

Together, these sets allow us to isolate the effects of domain proximity, model architecture, and cross-domain generalization in a controlled manner. For more details, refer \url{https://github.com/sa4s-serc/CALM}.




\subsection{Experimental Candidates}

To evaluate the efficacy of our proposed system, we consider two broad categories of experimental candidates: baselines that use a single Large Language Model (LLM), and various ablations of our approach that isolate key design components. This section details both the baseline setup and the systematic variations of the proposed approach for robust evaluation.

\subsubsection{Baseline: Single LLM Configuration}

As a reference point, we evaluate a baseline system employing a single LLM. To ensure fair comparison in terms of computational resources and memory footprint, we select LLMs whose peak memory consumption closely matches the cumulative memory usage of our deployed set of SLMs. We exclude API-based LLMs as their opaque execution environment introduces ambiguity making fair benchmarking of latency and energy consumption infeasible. The selected models used for baseline evaluation include:
\textbf{Deepseek-MOE 16B}, \textbf{Llama-2 13B}, \textbf{Qwen2.5-14B} with fp16 precision.
These baseline allows us to assess the gains in performance, resource efficiency, and QoS control achieved by transitioning from a single LLM setup to a modular, SLM-driven system.

\begin{table*}[]
\renewcommand{\arraystretch}{1.3}
\caption{Performance Comparison of \approach variants with Baseline LLMs on Set1}
\resizebox{0.95\textwidth}{!}{
\begin{tabular}{|lcccccccccl|}
\hline
\multicolumn{1}{|c|}{}                                                                                              & \multicolumn{3}{c|}{\textbf{Latency (sec)}}                                                                                                      & \multicolumn{3}{c|}{\textbf{Confidence Score}}                                                                                                 & \multicolumn{3}{c|}{\textbf{Energy Consumption(kJ)}}                                                                                           & \multicolumn{1}{c|}{}                                                                                           \\ \cline{2-10}
\multicolumn{1}{|c|}{\multirow{-2}{*}{\textbf{\begin{tabular}[c]{@{}c@{}}Experimental \\ Candidates\end{tabular}}}} & \multicolumn{1}{c|}{\textit{$40 \mid 256$}} & \multicolumn{1}{c|}{\textit{$20 \mid 256$}} & \multicolumn{1}{c|}{\textit{$20 \mid 512$}}          & \multicolumn{1}{c|}{\textit{$40 \mid 256$}} & \multicolumn{1}{c|}{\textit{$20 \mid 256$}} & \multicolumn{1}{c|}{\textit{$20 \mid 512$}}        & \multicolumn{1}{c|}{\textit{$40 \mid 256$}} & \multicolumn{1}{c|}{\textit{$20 \mid 256$}} & \multicolumn{1}{c|}{\textit{$20 \mid 512$}}        & \multicolumn{1}{c|}{\multirow{-2}{*}{\textbf{\begin{tabular}[c]{@{}c@{}}Peak VRAM\\ Memory Used\end{tabular}}}} \\ \hline
\multicolumn{1}{|l|}{Llama-2 13B}                                                                                   & 1277.92             & 1644.41                                     & \multicolumn{1}{c|}{3791.36}                         & \cellcolor[HTML]{F5B9B9}3.124               & \cellcolor[HTML]{F5B9B9}3.142               & \multicolumn{1}{c|}{\cellcolor[HTML]{F5B9B9}3.064} & \cellcolor[HTML]{F5B9B9}2.598               & \cellcolor[HTML]{F5B9B9}2.604               & \multicolumn{1}{c|}{\cellcolor[HTML]{F5B9B9}5.169} & $\sim$28.1 GB                                                                                                   \\
\multicolumn{1}{|l|}{Deepseek-MOE 16B}                                                                              & \cellcolor[HTML]{F5B9B9}1523.32             & \cellcolor[HTML]{F5B9B9}2077.91             & \multicolumn{1}{c|}{\cellcolor[HTML]{F5B9B9}4285.05} & 3.420                                       & 3.408                                       & \multicolumn{1}{c|}{\cellcolor[HTML]{F5B9B9}3.378} & 1.091               & 1.057               & \multicolumn{1}{c|}{\cellcolor[HTML]{CFFFCA}1.908} & \cellcolor[HTML]{F5B9B9}$\sim$32.7 GB                                                                           \\
\multicolumn{1}{|l|}{Qwen2.5-14B}                                                                                   & 1236.67             & 1433.63                                     & \multicolumn{1}{c|}{3749.74}                         & 3.443                                       & \cellcolor[HTML]{F5B9B9}3.273               & \multicolumn{1}{c|}{3.772}                         & \cellcolor[HTML]{F5B9B9}2.822               & \cellcolor[HTML]{F5B9B9}2.807               & \multicolumn{1}{c|}{\cellcolor[HTML]{F5B9B9}5.148} & $\sim$30.2 GB                                                                                                   \\
\multicolumn{1}{|l|}{CALM (Parallel)}                                                                          & 1206.53                                     & 1848.97                                     & \multicolumn{1}{c|}{\cellcolor[HTML]{F5B9B9}4417.05}                         & 3.952               & 3.952               & \multicolumn{1}{c|}{3.914}                         & 0.947                                       & 1.218                                       & \multicolumn{1}{c|}{1.959}                         & \cellcolor[HTML]{F5B9B9}$\sim$32.5 GB                                                                           \\
\multicolumn{1}{|l|}{CALM (FIFO)}                                                                              & \cellcolor[HTML]{CFFFCA}761.13              & \cellcolor[HTML]{CFFFCA}975.80              & \multicolumn{1}{c|}{\cellcolor[HTML]{CFFFCA}2354.84} & \cellcolor[HTML]{CFFFCA}4.006               & \cellcolor[HTML]{CFFFCA}4.020               & \multicolumn{1}{c|}{3.950} & 1.023                                       & 0.964                                       & \multicolumn{1}{c|}{\cellcolor[HTML]{CFFFCA}1.905} & \cellcolor[HTML]{F5B9B9}$\sim$32.4 GB                                                                           \\
\multicolumn{1}{|l|}{CALM (STCF)}                                                                              & \cellcolor[HTML]{CFFFCA}686.11              & \cellcolor[HTML]{CFFFCA}1011.37             & \multicolumn{1}{c|}{\cellcolor[HTML]{CFFFCA}2455.86} & \cellcolor[HTML]{CFFFCA}4.106               & 3.964               & \multicolumn{1}{c|}{3.978} & 1.049                                       & \cellcolor[HTML]{CFFFCA}0.986               & \multicolumn{1}{c|}{1.962}                         & \cellcolor[HTML]{F5B9B9}$\sim$32.5 GB                                                                           \\ \hline
\multicolumn{11}{|c|}{\textbf{With Cache Module}}                                                                                                                                                                                                                                                                                                                                                                                                                                                                                                                                                                                                                                          \\ \hline
\multicolumn{1}{|l|}{CALM (CM, cache\_size = 1) (FIFO)}                                                        & \cellcolor[HTML]{F5B9B9}1703.64             & \cellcolor[HTML]{F5B9B9}2314.08                                     & \multicolumn{1}{c|}{3673.02}                         & 3.943               & 3.958               & \multicolumn{1}{c|}{3.958} & \cellcolor[HTML]{CFFFCA}0.994                                       & \cellcolor[HTML]{CFFFCA}0.952                                       & \multicolumn{1}{c|}{\cellcolor[HTML]{CFFFCA}1.918} & \cellcolor[HTML]{CFFFCA}$\sim$7.9 GB                                                                            \\
\multicolumn{1}{|l|}{CALM (CM, cache\_size = 3) (FIFO)}                                                        & 1358.64             & 1863.26                                     & \multicolumn{1}{c|}{3313.61} & 3.932                                       & 3.970               & \multicolumn{1}{c|}{3.884}                         & \cellcolor[HTML]{CFFFCA}0.942               & \cellcolor[HTML]{CFFFCA}0.934               & \multicolumn{1}{c|}{2.003}                         & \cellcolor[HTML]{CFFFCA}$\sim$18.8 GB                                                                           \\
\multicolumn{1}{|l|}{CALM (CM, cache\_size = 3) (STCF)}                                                        & \cellcolor[HTML]{CFFFCA}982.15              & \cellcolor[HTML]{CFFFCA}1170.64             & \multicolumn{1}{c|}{\cellcolor[HTML]{CFFFCA}2488.80} & 3.952               & 3.936                                       & \multicolumn{1}{c|}{3.890}                         & 1.003                                       & 1.029                                       & \multicolumn{1}{c|}{2.076}                         & \cellcolor[HTML]{CFFFCA}$\sim$18.7 GB                                                                           \\ \hline
\end{tabular}
}
\captionsetup{justification=justified}
\caption*{\footnotesize Caption: \textnormal{Table shows performance comparison averaged over 500 queries using Set 1. Here, high latency is due to queued requests on a shared single-GPU for all SLMs under heavy load. Despite high absolute times, consistent setup across candidates allows for valid relative comparisons.}}
\vspace{-10 pt}
\label{Table1}
\end{table*}

\subsubsection{Router \& Scheduler}

The router refers to the algorithm used to route user queries to the most relevant SLM based on model descriptions.
We evaluate multiple similarity-based routing methods \textbf{Cosine Similarity},  \textbf{MaxSim}~\cite{chan2008maxsim}, and \textbf{CrossEncoder}~\cite{reimers2019sentencebert}—using three metrics: \textit{decision time} (time to generate routing decision), \textit{confidence score}, and \textit{decision accuracy}. \textcolor{black}{To compute the Decision Accuracy, we take the user queries from each of the datasets mentioned in Section 4.1.2 and run the setup to route the queries only based on the static similarity scores. Decision accuracy is then computed as the percentage of queries routed to the correct domain model. The remaining metrics are also calculated based on the definitions.}
As shown in Table~\ref{tab:router}, \textbf{Cosine Similarity} consistently achieves the best balance of high decision accuracy and low decision time, making it the default routing method in our experiments.

The scheduler is the algorithm which manages request execution across the model fleet. We evaluate three strategies: First-In-First-Out (FIFO), Shortest Time to Completion First (STCF), and Parallel Routing. Due to single-GPU constraints, parallel routing incurs significant overhead from context switching. As shown in Table~\ref{Table1}, FIFO and STCF yield comparable latency, but to avoid starvation issues with STCF, we adopt FIFO for future experiments.

\begin{table}[H]
\centering
\renewcommand{\arraystretch}{1.2}
\caption{Comparison of Routing Techniques}
\small

\resizebox{0.8\columnwidth}{!}{%
\begin{tabular}{|l|c|c|c|}
\hline
\textbf{Router} &
\textbf{\begin{tabular}[c]{@{}c@{}}Decision \\ Time (ms)\end{tabular}} &
\textbf{\begin{tabular}[c]{@{}c@{}}Confidence \\ Score (/5)\end{tabular}} &
\textbf{\begin{tabular}[c]{@{}c@{}}Decision \\ Acc. (\%)\end{tabular}} \\
\hline
Cosine Similarity      & 38.434 & 3.623 & 77.2 \\
MaxSim                 & 97.557 & 3.615 & 66.4 \\
CrossEncoder Relevancy & 19.185 & 3.563 & 61.2 \\
\hline
\end{tabular}
}

\vspace{1mm}
\captionsetup{justification=justified, skip=3pt}
\caption*{\footnotesize
Table shows different routing techniques to compute static\_similarity.
Evaluation metrics are averaged over 500 queries using Set~3.
Lower decision time and higher accuracy indicate more efficient and relevant routing.
}

\vspace{-8pt}
\label{tab:router}
\end{table}

\subsubsection{\approach: Configurations}

We perform experimentatal studies to analyze the contribution of individual components in \approach aproach. The base approach consists of routing user queries to the most relevant SLM based on static similarity scores. We then incrementally integrate caching and self-adaptive routing to assess their individual and combined impacts.

     \textbf{Base Approach:} To the best of our knowledge, very few self-adaptive baselines exist for ML systems, such as SWITCH \cite{marda2024switch}. We therefore adapt SWITCH’s ML-based routing mechanism and replace it with a static domain-based routing to forward queries to specialized models, without caching or adaptive mechanisms as a baseline.\textbf{6 SLMs are deployed on a shared GPU}. It isolates the benefits of using SLMs with domain-aware routing over a single LLM. We refer to this approach as \approach throughout the paper, with specific configuration details provided in parentheses where applicable. 

     \textbf{Base Approach + Model Caching:} This setup introduces a caching mechanism to the base system, helping reduce cold starts and memory consumption. We analyze the effect of varying cache sizes (1–3 ie, maximum of 1-3 models can be in the memory) on cache hit rates and system responsiveness, under fixed resource constraints. We denote this configuration as \approach (CM), where CM indicates the use of the Caching Module.

     \textbf{Base Approach + Self-Adaptive Routing:} In this variant, we augment the base routing system with dynamic adaptation based on QoS metrics, without employing the caching module. The system adjusts model selection strategies using a \textbf{tunable control parameter $\lambda$}, allowing service providers to prioritize metrics like latency or confidence. We evaluate 3 prioritization strategies:\\
    \textbf{Pro-Latency Routing} where the model selection score is given as:
    \[
    \text{score} = (1 - \text{similarity\_score}) + \lambda_{lat} * \hat{L}
    \]
    Here, similarity\_score represents the semantic alignment between the query and the model’s task description, $\hat{L}$ is the normalized average inference times (scaled to [0, 1]) and $\lambda_{lat} \in \{0.1, 0.2, 0.3\}$. The model with the lowest score is selected. \\
     \textbf{Pro-Energy Routing} where the model selection score is given as:
    \[
    \text{score} = (1 - \text{similarity\_score}) + \lambda_{energy} * \hat{E}
    \]
    Here, $\hat{L}$ is the normalized average energy consumptions (scaled to [0, 1]) and $\lambda_{energy} \in \{0.1, 0.15, 0.2\}$. \\
    \textbf{Pro-Confidence Routing} where selection score is defined as:
    \[
    \text{score} = \text{similarity\_score} + \lambda_{conf} * \hat{C}
    \]
    Here, $\hat{C}$ is the normalized average confidence (scaled to [0, 1] from raw values ranging from 0 to 5), and $\lambda_{conf} \in \{0.1, 0.2, 0.3\}$. The model with the highest score is selected.\\
    \textbf{Pro-Cache Routing} where $\lambda_{cache}$ is added to the score for models already present in the cache module to optimize cache hits:
    \[
    \text{score} = \text{similarity\_score} + \lambda_{cache} 
    \]
    Here $\lambda_{cache} \in \{0.02, 0.03, 0.05\}$ is the bias control parameter. The model with the highest score is selected.
We refer these approachs as \approach($\lambda = k$) where k is the value of the control parameter. \color{black} For cold-start scenarios with no historical data, we initialize latency and energy metrics to zero and confidence to a 5, ensuring that early routing decisions are driven primarily by similarity scores and that all models receive fair initial exploration. \color{black}
    \vspace{3pt}

     \textbf{Combined Approach:} Finally, we evaluate the full system that integrates domain-based routing, model caching, and self-adaptive routing. This configuration allows us to assess the holistic benefits of our architecture in terms of performance, memory efficiency, and QoS control and is the final point of comparison with  the LLM baseline. We refer these as \approach(CM, $\lambda = k$).

For all experiments, we use \textbf{FIFO scheduling} and \textbf{Cosine Similarity} as the domain router unless explicitly mentioned in parentheses. Additionally, the Caching Module employs a \textbf{Least Recently Used (LRU)} eviction policy to dynamically manage memory by removing underutilized models over time. LRU prioritizes frequently used models by retaining the most recently accessed ones in memory, which aligns with the intuition that models used more often are more likely to be needed again soon. Similarly, for all experiments, if a Caching Module (CM) is present, the cache size is set to \textbf{3} unless explicitly mentioned in parentheses.

\subsection{Experimental Setup}

\subsubsection{Hardware Configuration}

\textcolor{black}{All model queries were routed over TCP across compute nodes to emulate a realistic networked serving setup. The experimental testbed consisted of three compute nodes: (i) one inference node, equipped with a single NVIDIA RTX A6000 (48 GB VRAM) and a 12-core CPU, hosting all deployed models (with active models resident in VRAM and inactive models stored on disk); (ii) one routing node, dedicated to running SORA, which listens for incoming client requests and performs model selection and request forwarding; and (iii) one workload generation node, used to simulate concurrent client requests. Communication between all nodes occurred over TCP to reflect a realistic serving environment.}

\subsubsection{Evaluation}

We track four key metrics: latency, energy consumption, response confidence, and peak VRAM usage, using standardized procedures. \textbf{Latency (in seconds)} is measured as the time from query arrival to response. \textbf{Energy (in kJ)} is estimated via \textcolor{black}{VRAM} usage recorded before and after processing \textcolor{black}{the user request} using NVIDIA Management Library (NVML). \textbf{Confidence scores (1–5 Likert scale)} are assigned via GPT-4o-based evaluation or real-time client feedback. To simulate feedback, GPT-4o scores outputs on helpfulness, relevance, and factuality, serving as a proxy for user satisfaction, following~\cite{gu2025surveyllmasajudge}. All metrics are aggregated over a rolling window of the last \textbf{100} requests. Peak \textcolor{black}{VRAM} usage is monitored using NVML during each run.

\section{Results}

\textbf{RQ 1.1}: \space \space \textbf{How does the system perform across key metrics—latency,
confidence score, energy consumption, when compared to
baseline LLMs?}

\begin{figure}[htbp]
\vspace{-10pt}
\caption{Variation of Latency across Candidates }
\centerline{\includegraphics[width=0.8\columnwidth]{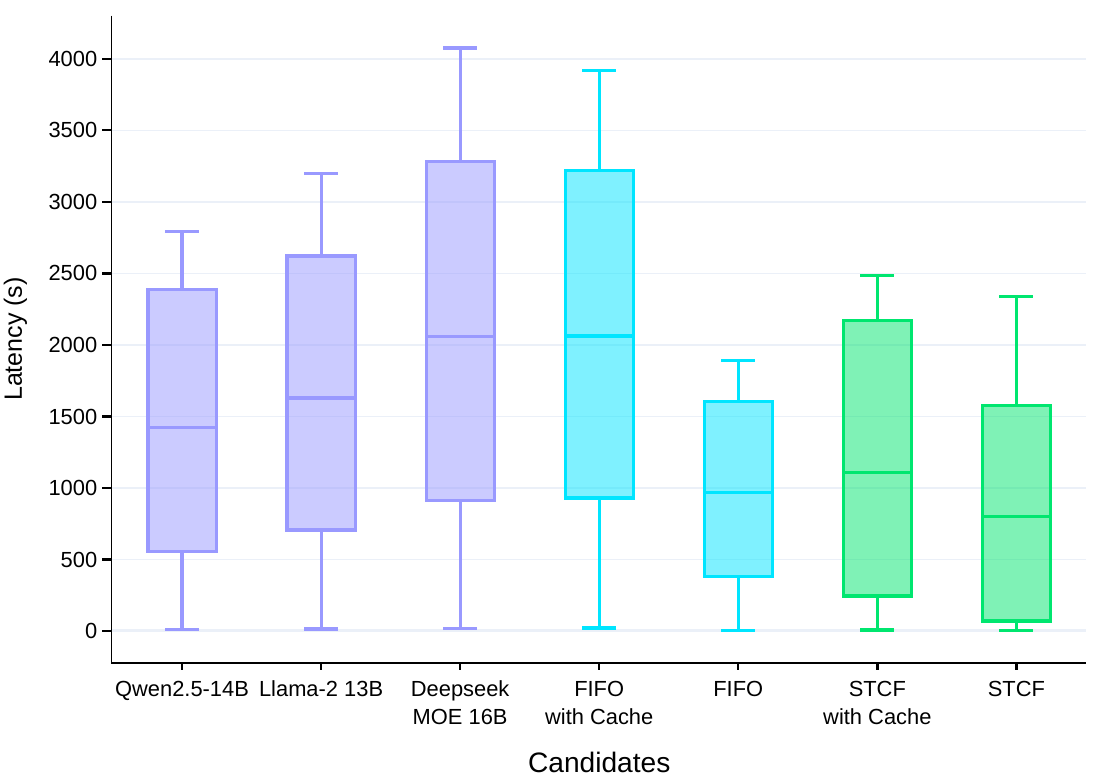}}
\label{fig}
\vspace{-5pt}
\end{figure}

Table~\ref{Table1} shows that \approach variants consistently outperform baseline LLMs across latency, confidence, energy, and memory, demonstrating practical viability under diverse workloads. For \textbf{latency}, FIFO and STCF scheduling reduce average latency by over \textbf{40--50\%} compared to baselines, due to the lightweight nature and efficiency of SLMs. While \approach (STCF) achieves slightly lower average latency than \approach (FIFO), it incurs higher variance (Figure~\ref{fig}), suggesting frequent starvations for a subset of requests. This makes FIFO scheduling preferable where predictability and fairness matters, such as in user-facing systems. Both variants show significantly lower latency variance than baseline LLMs, indicating more stable real-time behavior under load. \textbf{Confidence scores} for all 3 scheduling strategies are \textbf{15--30\%} higher than baseline LLMs, reflecting better task alignment through specialization, which is useful in scenarios requiring accurate, factually correct responses particularly for medical or legal domains. \textbf{Energy consumption} offers the sharpest gains: \approach (FIFO/STCF) uses \textbf{50--60\%} less energy than full-scale LLMs, while remaining competitive with sparsely activated MoE models like Deepseek-MOE. This positions \approach as a more efficient option for constrained or cost-sensitive environments. Even cache-enabled variants of \approach outperform baselines in latency across all workloads. Despite cold starts, which occur when a queried model isn't in memory. \approach remains faster, making it responsive even under limited memory availability.

\begin{figure}[htbp]
\vspace{-10pt}

\caption{\textcolor{black}{Tail latency analysis}}
\centerline{\includegraphics[width=0.8\columnwidth]{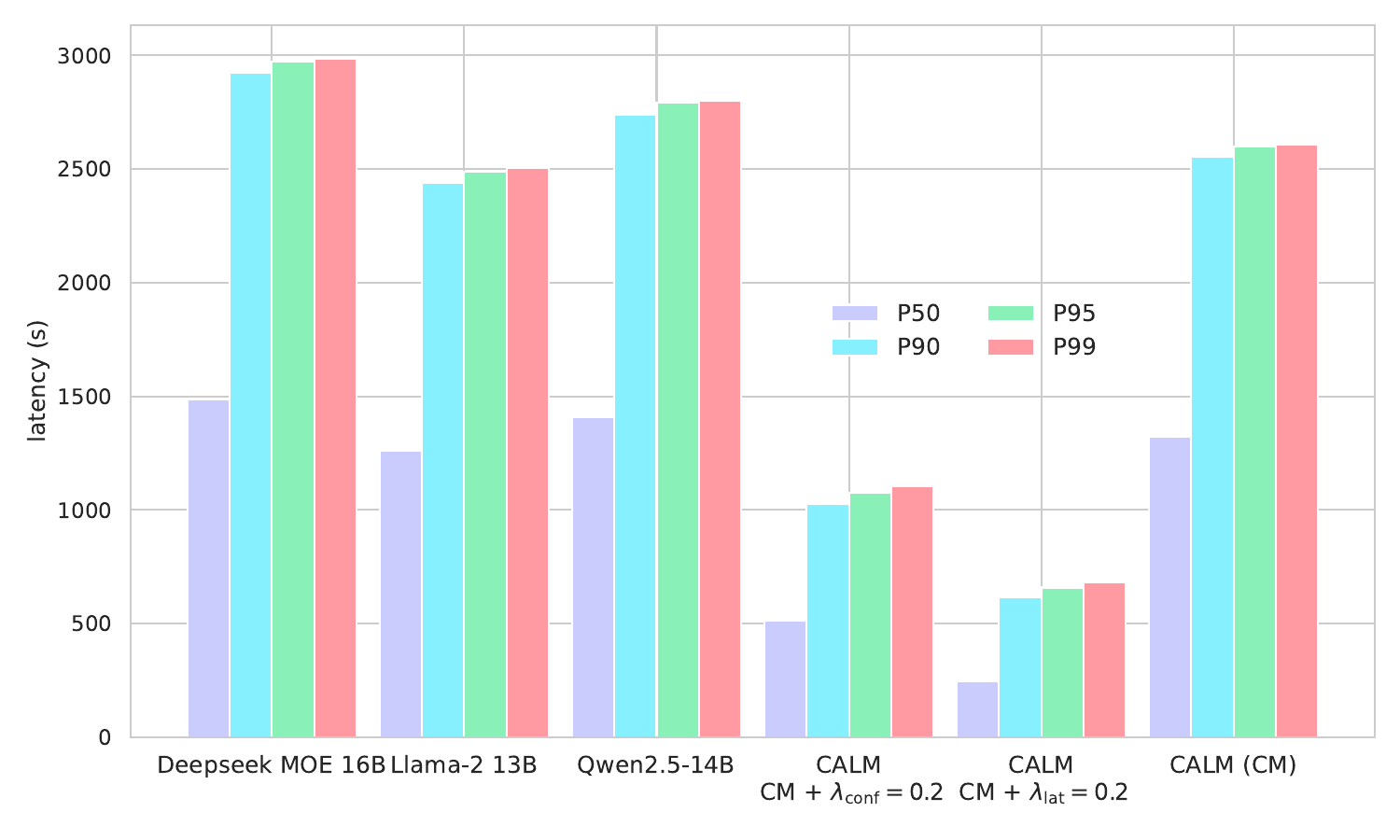}}
\label{fig:tail}
\vspace{-10pt}
\vspace{1mm}
\captionsetup{justification=justified, skip = 3 pt}

\end{figure}
\textcolor{black}{While average latency provides a coarse summary of system behavior, it can obscure request-level effects under contention. We therefore additionally examine tail latencies (p90--p99), which more accurately capture worst-case user experience. As shown in Figure~\ref{fig:tail}, all approaches exhibit a similar tail-latency structure, reflecting the shared single-GPU deployment and serialized inference execution, which inherently amplifies queuing delays at higher percentiles. Importantly, however, all \approach variants consistently achieve lower tail latencies than baseline LLMs across p90--p99, indicating that dynamic routing to specialized lightweight models does not exacerbate worst-case latency, despite per-query model selection and occasional cold starts. Although tail latencies remain elevated in absolute terms, this behavior is attributable to the constrained hardware setting rather than the routing or scheduling policy itself, making the comparison fair across systems.}

\begin{figure}[htbp]
\vspace{-10pt}

\caption{\textcolor{black}{Average Inference Times of SLMs across examples}}
\centerline{\includegraphics[width=0.85\columnwidth]{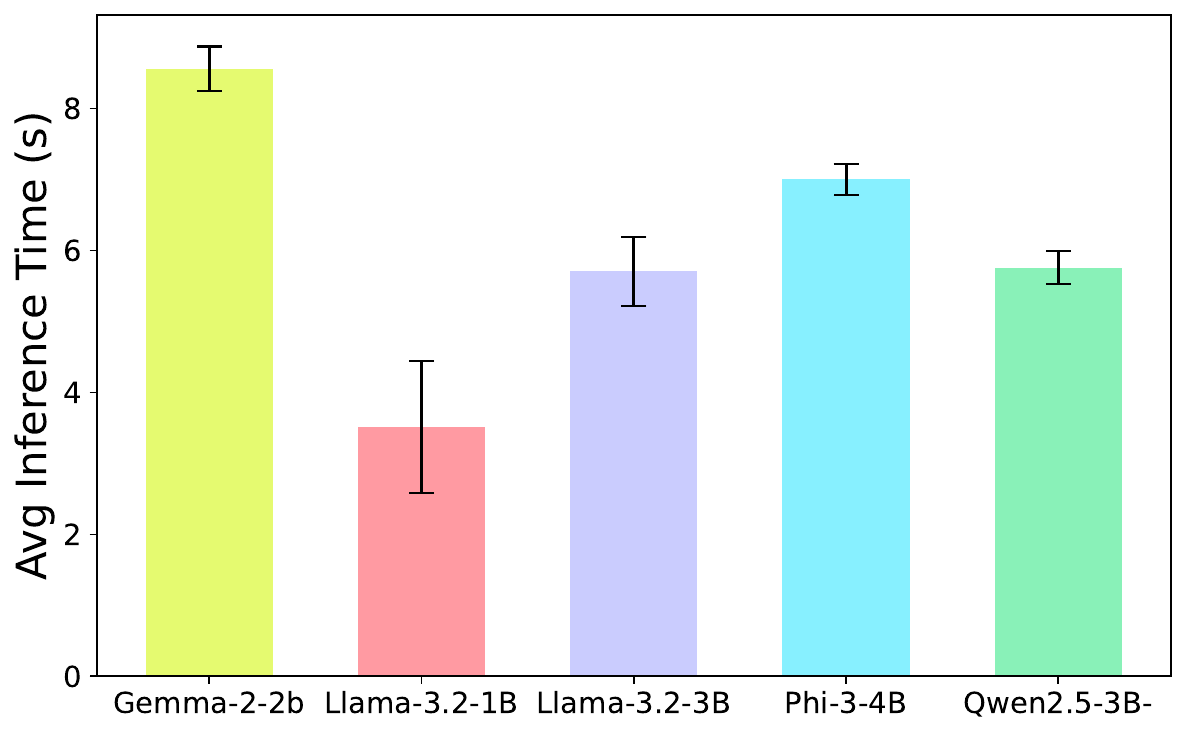}}
\label{fig:avg_ing}
\vspace{-10pt}
\vspace{1mm}
\captionsetup{justification=justified, skip = 3 pt}

\end{figure}

\begin{tcolorbox}[summarybox]
Significant latency and energy reductions (40-50\%) makes \approach highly suitable for real-time systems with tight performance and power budgets. Higher confidence gains (15-30\%) through domain-specialization supports adoption in narrowly scoped, domain-specific application settings that require extensive specialized knowledge and impose stringent correctness requirements
\end{tcolorbox}

\textbf{RQ 1.2}: \space \space \textbf{How does the cache module impact memory utilization while maintaining QoS?}

Table~\ref{Table1} shows that adding a caching layer to \approach reduces VRAM usage from \textbf{32~GB to as low as 8~GB}, without substantial degradation in latency, confidence score, or energy efficiency as compared to baseline LLMs. Even with limited in-memory model availability, the system maintains stable and responsive performance, enabling memory-efficient deployment such as edge GPUs or shared inference clusters without compromising output quality. With a moderate cache size of 3,  incurs only a \textbf{20-25\%} increase in latency (owing to the time taken for cold starts when a model has to be loaded in the memory) relative to its non-cached variant but reduces \textcolor{black}{VRAM} consumption by \textbf{35-45\%}, while confidence scores and energy consumption remain comparable to the best-performing \approach configurations. This reflects a favorable trade-off between memory savings and runtime performance. Among cache strategies, STCF achieves lower latency than FIFO by minimizing cache switching, prioritizing queries for the fastest model before eviction. FIFO, while fairer, incurs higher latency due to more frequent model replacements. 

\begin{tcolorbox}[summarybox]
Cache Module of moderate size in \approach enables deployment on memory-constrained environments by cutting VRAM usage around 35-45\% with minimal impact on quality and energy but introduces a latency overhead due to cold starts.
\end{tcolorbox}

\begin{table}[]
\renewcommand{\arraystretch}{1.1}
\caption{Self-Adaptive Optimization of Latency, Confidence and Energy Consumption across Control parameters (Set2)}
\tiny
\resizebox{0.9\columnwidth}{!}{
\begin{tabular}{|cccc|}
\hline
\multicolumn{1}{|c|}{\textbf{\begin{tabular}[c]{@{}c@{}}Control\\ Parameter\end{tabular}}} & \multicolumn{1}{c|}{\textbf{\begin{tabular}[c]{@{}c@{}}Latency\\ (sec)\end{tabular}}} & \multicolumn{1}{c|}{\textbf{\begin{tabular}[c]{@{}c@{}}Confidence\\ Score (/5)\end{tabular}}} & \textbf{\begin{tabular}[c]{@{}c@{}}Energy\\ Consumption (kJ)\end{tabular}} \\ \hline
\multicolumn{4}{|c|}{\textbf{Pro-Latency Routing}}                                                                                                                                                                                                                                                                                                         \\ \hline
\multicolumn{1}{|c|}{$\lambda_{\scalebox{0.7}{$\text{}$}}=0$}                                                                    & \multicolumn{1}{c|}{766.480}                                  & \multicolumn{1}{c|}{\cellcolor[HTML]{CFFFCA}3.892}                                            & 0.8655                                                                \\
\multicolumn{1}{|c|}{$\lambda_{\scalebox{0.7}{$\text{lat}$}}=0.1$}                                                                  & \multicolumn{1}{c|}{665.920}                                  & \multicolumn{1}{c|}{3.750}                                            & 0.7366                                        \\
\multicolumn{1}{|c|}{$\lambda_{\scalebox{0.7}{$\text{lat}$}}=0.2$}                                                                  & \multicolumn{1}{c|}{\cellcolor[HTML]{CFFFCA}508.965}                                  & \multicolumn{1}{c|}{3.490}                                                                    & 0.6473                                        \\
\multicolumn{1}{|c|}{$\lambda_{\scalebox{0.7}{$\text{lat}$}}=0.3$}                                                                  & \multicolumn{1}{c|}{\cellcolor[HTML]{CFFFCA}430.929}                                  & \multicolumn{1}{c|}{\cellcolor[HTML]{F5B9B9}3.248}                                            & \cellcolor[HTML]{CFFFCA}0.5474                                        \\ \hline
\multicolumn{4}{|c|}{\textbf{Pro-Latency Routing + Cache Module}}                                                                                                                                                                                                                                                                                          \\ \hline
\multicolumn{1}{|c|}{$\lambda_{\scalebox{0.7}{$\text{}$}}=0$}                                                                    & \multicolumn{1}{c|}{\cellcolor[HTML]{F5B9B9}1400.653}                                 & \multicolumn{1}{c|}{\cellcolor[HTML]{CFFFCA}3.908}                                            & 1.0430                                                                \\
\multicolumn{1}{|c|}{$\lambda_{\scalebox{0.7}{$\text{lat}$}}=0.1$}                                                                  & \multicolumn{1}{c|}{\cellcolor[HTML]{F5B9B9}1246.508}                                 & \multicolumn{1}{c|}{3.728}                                            & 0.7198                                        \\
\multicolumn{1}{|c|}{$\lambda_{\scalebox{0.7}{$\text{lat}$}}=0.2$}                                                                  & \multicolumn{1}{c|}{876.860}                                                          & \multicolumn{1}{c|}{3.452}                                                                    & 0.6094                                        \\
\multicolumn{1}{|c|}{$\lambda_{\scalebox{0.7}{$\text{lat}$}}=0.3$}                                                                  & \multicolumn{1}{c|}{\cellcolor[HTML]{CFFFCA}550.223}                                  & \multicolumn{1}{c|}{\cellcolor[HTML]{F5B9B9}3.186}                                            & \cellcolor[HTML]{CFFFCA}0.5247                                        \\ \hline
\multicolumn{4}{|c|}{\textbf{Pro-Confidence Routing}}                                                                                                                                                                                                                                                                                                      \\ \hline
\multicolumn{1}{|c|}{$\lambda_{\scalebox{0.7}{$\text{}$}}=0$}                                                                    & \multicolumn{1}{c|}{766.480}                                  & \multicolumn{1}{c|}{3.892}                                            & 0.8655                                                                \\
\multicolumn{1}{|c|}{$\lambda_{\scalebox{0.7}{$\text{conf}$}}=0.1$}                                                                  & \multicolumn{1}{c|}{844.990}                                                          & \multicolumn{1}{c|}{3.950}                                            & 0.9016                                                                \\
\multicolumn{1}{|c|}{$\lambda_{\scalebox{0.7}{$\text{conf}$}}=0.2$}                                                                  & \multicolumn{1}{c|}{803.976}                                                          & \multicolumn{1}{c|}{\cellcolor[HTML]{CFFFCA}4.076}                                            & 0.8678                                                                \\
\multicolumn{1}{|c|}{$\lambda_{\scalebox{0.7}{$\text{conf}$}}=0.3$}                                                                  & \multicolumn{1}{c|}{932.265}                                                          & \multicolumn{1}{c|}{\cellcolor[HTML]{CFFFCA}4.228}                                            & 0.9701                                                                \\ \hline
\multicolumn{4}{|c|}{\textbf{Pro-Confidence Routing + Cache Module}}                                                                                                                                                                                                                                                                                       \\ \hline
\multicolumn{1}{|c|}{$\lambda_{\scalebox{0.7}{$\text{}$}}=0$}                                                                    & \multicolumn{1}{c|}{\cellcolor[HTML]{F5B9B9}1400.653}                                 & \multicolumn{1}{c|}{3.908}                                            & 1.0430                                                                \\
\multicolumn{1}{|c|}{$\lambda_{\scalebox{0.7}{$\text{conf}$}}=0.1$}                                                                  & \multicolumn{1}{c|}{\cellcolor[HTML]{F5B9B9}1376.157}                                 & \multicolumn{1}{c|}{3.938}                                            & 1.0857                                                                \\
\multicolumn{1}{|c|}{$\lambda_{\scalebox{0.7}{$\text{conf}$}}=0.2$}                                                                  & \multicolumn{1}{c|}{1079.684}                                                         & \multicolumn{1}{c|}{3.958}                                            & 1.021                                                                 \\
\multicolumn{1}{|c|}{$\lambda_{\scalebox{0.7}{$\text{conf}$}}=0.3$}                                                                  & \multicolumn{1}{c|}{992.878}                                                          & \multicolumn{1}{c|}{\cellcolor[HTML]{CFFFCA}4.268}                                            & 1.1843                                        \\ \hline
\multicolumn{4}{|c|}{\textbf{Pro-Energy Routing}}                                                                                                                                                                                                                                                                                       \\ \hline
\multicolumn{1}{|c|}{$\lambda_{\scalebox{0.7}{$\text{}$}}=0$}                                                                    & \multicolumn{1}{c|}{766.480 }                                 & \multicolumn{1}{c|}{3.892 }                                            & 0.8655                                                               \\
\multicolumn{1}{|c|}{$\lambda_{\scalebox{0.7}{$\text{energy}$}}=0.1$}                                                                  & \multicolumn{1}{c|}{692.294}                                 & \multicolumn{1}{c|}{3.696}                                            & 0.8033                                                                \\
\multicolumn{1}{|c|}{$\lambda_{\scalebox{0.7}{$\text{energy}$}}=0.15$}                                                                  & \multicolumn{1}{c|}{502.357}                                                         & \multicolumn{1}{c|}{3.604}                                            & 0.6963                                                                \\
\multicolumn{1}{|c|}{$\lambda_{\scalebox{0.7}{$\text{energy}$}}=0.2$}                                                                  & \multicolumn{1}{c|}{\cellcolor[HTML]{CFFFCA}476.496}                                                          & \multicolumn{1}{c|}{3.536}                                            & \cellcolor[HTML]{CFFFCA}0.5563                                        \\ \hline
\multicolumn{4}{|c|}{\textbf{Pro-Energy Routing + Cache Module}}                                                                                                                                                                                                                                                                                       \\ \hline
\multicolumn{1}{|c|}{$\lambda_{\scalebox{0.7}{$\text{}$}}=0$}                                                                    & \multicolumn{1}{c|}{\cellcolor[HTML]{F5B9B9}1400.653}                                 & \multicolumn{1}{c|}{3.908}                                            & 1.0430                                                                \\
\multicolumn{1}{|c|}{$\lambda_{\scalebox{0.7}{$\text{energy}$}}=0.1$}                                                                  & \multicolumn{1}{c|}{724.748}                                 & \multicolumn{1}{c|}{3.811}                                            & 0.6672                                                               \\
\multicolumn{1}{|c|}{$\lambda_{\scalebox{0.7}{$\text{energy}$}}=0.15$}                                                                  & \multicolumn{1}{c|}{565.042}                                                         & \multicolumn{1}{c|}{3.748}                                            & \cellcolor[HTML]{CFFFCA}0.5943                                                                 \\
\multicolumn{1}{|c|}{$\lambda_{\scalebox{0.7}{$\text{energy}$}}=0.2$}                                                                  & \multicolumn{1}{c|}{529.249}                                                          & \multicolumn{1}{c|}{3.738}                                            & \cellcolor[HTML]{CFFFCA}0.5674                                        \\ \hline
\end{tabular}
\vspace{-9pt}
}
\label{tab:self-adaptive}
\end{table}
\begin{table}[]
\renewcommand{\arraystretch}{1.2}
\caption{Pro-Cache Routing Performance(size=3) (Set 2)}
\resizebox{0.9\columnwidth}{!}{
\begin{tabular}{|l|c|c|c|c|}
\hline
\multicolumn{1}{|c|}{\textbf{\begin{tabular}[c]{@{}c@{}}Control\\ Parameter\end{tabular}}} & \textbf{\begin{tabular}[c]{@{}c@{}}Latency\\ (sec)\end{tabular}} & \textbf{\begin{tabular}[c]{@{}c@{}}Confidence\\ Score (/5)\end{tabular}} & \textbf{\begin{tabular}[c]{@{}c@{}}Energy\\ Consumption (kJ)\end{tabular}} & \multicolumn{1}{l|}{\textbf{\begin{tabular}[c]{@{}l@{}}Cache\\ Hit (\%)\end{tabular}}} \\ \hline
$\lambda_{\scalebox{0.7}{$\text{}$}}=0$                    & \cellcolor[HTML]{F5B9B9}1400.653                                                         & 3.908                                                                    & \cellcolor[HTML]{F5B9B9}1.0430                                             & \cellcolor[HTML]{F5B9B9}57.6                                                           \\
$\lambda_{\scalebox{0.7}{$\text{cache}$}}=0.02$                  & 1285.786                                                         & \cellcolor[HTML]{CFFFCA}3.95                                                                     & 0.8685                                                                     & 68.0                                                           \\
$\lambda_{\scalebox{0.7}{$\text{cache}$}}=0.03$                     & 1136.723                                                         & 3.876                                                                    & \cellcolor[HTML]{CFFFCA}0.8378                                                                     & 71.6                                                           \\
$\lambda_{\scalebox{0.7}{$\text{cache}$}}=0.05$                     & \cellcolor[HTML]{CFFFCA}1120.493                                 & \cellcolor[HTML]{CFFFCA}3.97                                             & 0.8944                                                                     & \cellcolor[HTML]{CFFFCA}83.8                                                           \\ \hline
\end{tabular}
}
\label{tab:caching-adaptive}
\end{table}

\textbf{RQ 1.3}: \space \space \textbf{How can a self-adaptive MAPE-K loop improve Quality of Service (QoS) by dynamically selecting models by incorporating real-time performance metrics?}

Table~\ref{tab:self-adaptive} demonstrates that the self-adaptive framework enables fine-grained QoS control by dynamically adjusting performance trade-offs.
With Pro-Latency Routing in \approach, increasing the control parameter ($\lambda_{lat}$) steadily reduces latency, with upto \textbf{44\% improvement}, while maintaining acceptable confidence levels. The energy consumption also drops by about \textbf{35-40\%} as system prefers models with faster inference and fewer computations. This reflects the approach’s flexibility in adapting to real-time constraints through specialized routing. Even with Cache Module, the improvement trends persist, while memory usage remain consistent across $\lambda_{lat}$ , confirming the scalability of the approach under memory constraints. Pro-Energy shows effects similar to the Pro-Latency Routing but with a higher sensitivity towards $\lambda_{energy}$ while being robust to the Confidence Score. Conversely, in Pro-Confidence Routing, response quality improves steadily (upto 10\%) as $\lambda_{conf}$ increases, with moderate rises in latency and energy consumption. Surprisingly, in Pro-Confidence routing with Cache Module, latency drops (Table \ref{tab:self-adaptive}) as higher-confidence models are kept in cache, leading to more effective cache hits. Table~\ref{tab:caching-adaptive} shows that increasing ($\lambda_{cache}$) significantly boosts cache hit rates giving slight priority to cached models. This reduces latency owing to \textbf{lesser number of cold starts}, without any adverse effect on confidence or energy, validating the benefit of Pro-Cache Routing. 

\begin{tcolorbox}[summarybox] 
\vspace{-3pt}
Incorporation of self-adaptive framework enables fine-grained QoS control by dynamically routing queries based on desired trade-offs.
Tuning the control parameter yields up to 44\% lower latency or top confidence scores, with efficient energy and memory use. Cache-aware routing further boosts performance by improving hit rates and reducing load-evict time overhead.
\end{tcolorbox}
\begin{figure}[htbp]
\vspace{-10pt}

\caption{Latency \& Confidence Tradeoffs for Self-Adaptive }
\centerline{\includegraphics[width=0.9\columnwidth]{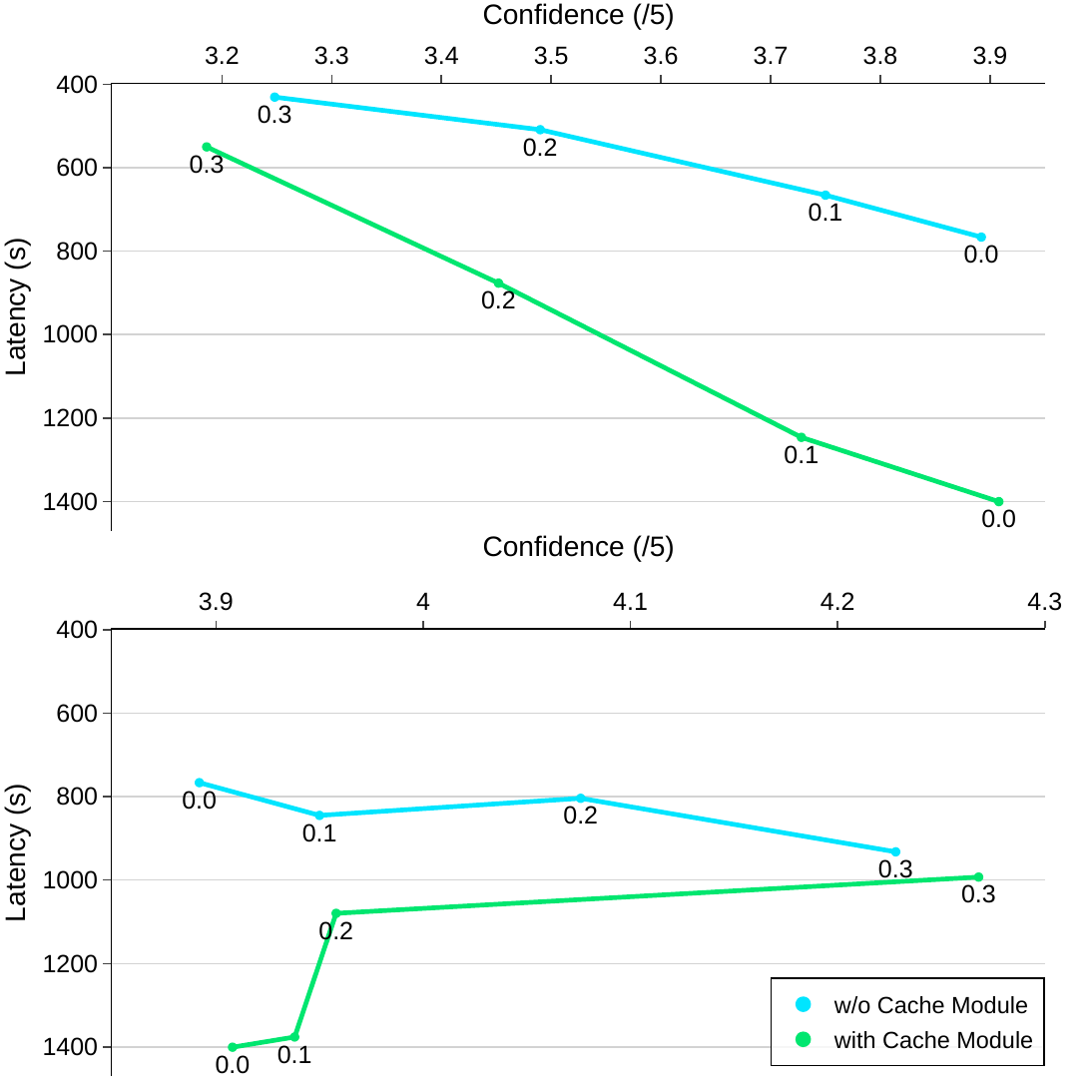}}
\label{fig:sa_tradeoff}
\vspace{1mm}
\captionsetup{justification=justified, skip = 3 pt}
\caption*{\footnotesize Caption: \textnormal{Extrapolation of Latency and Confidence Values tradeoff (with \& w/o Cache) variants for Pro-Latency (top) and Pro-Confidence (bottom) Routing. Points denote \(\lambda_{lat}\) and \(\lambda_{conf}\) respectively.}}
\vspace{-7 pt}
\end{figure}

\textbf{RQ2: How well does the system scale to different workloads
and generalize to diverse domains?} 
\subsubsection*{Scalability:}  
Table~\ref{Table1} highlights the workload scalability of the \approach framework, i.e., how performance changes under increasing incoming request rates. All variants, \approach (FIFO), \approach (STCF), and their cache-enhanced counterparts, show a predictable rise in latency and energy consumption as the workload increases (from 40 | 256 to 20 | 512), yet consistently outperform single LLM baselines. Compared to Llama-2 and Qwen2.5, \approach exhibits \textbf{15--25\% lower latency and 20--35\%} lower energy usage under load. With Deepseek-MOE as well, the improvements remain \textbf{consistent across workloads}. Confidence scores remain stable across configurations, since they depend on model response quality, not the overall load on a model, unless the routing policy also enforces prioritization for latency/confidence. From a scalability standpoint, the LRU caching mechanism further enhances system capacity by enabling \approach to abstract a pool of 6 models while \textbf{maintaining only 3 in memory} at any time, effectively scaling the number of supported models under a fixed \textcolor{black}{VRAM} budget with minimal latency overhead. These trends underscore how memory-aware design choices can unlock scalable deployment of specialized LLMs in real-world, resource-constrained settings.

\subsubsection*{Generalizability:}  

From Table~\ref{tab:self-adaptive}, with \(\lambda = 0\), we observe the \approach configurations for Set 2 models on the same domain as Table~\ref{Table1} (see Section 4.1.2). The results show that latency, confidence, and energy consumption consistently outperform the baseline LLMs. This confirms that the model selection process does not introduce bias, demonstrating that the approach is model-agnostic.
As shown in Table~\ref{tab:generalizability}, \approach achieves significantly lower latency and energy consumption even in inter-domain settings, improving both by \textbf{30--50\%} as observed in (Table~\ref{Table1}) as well, owing to the \approach approach and same workloads. However, this is achieved while maintaining confidence scores on par with larger LLMs, even in the presence of inter-domain challenges like \textbf{routing ambiguity} and sensitivity to \textbf{fine-tuning quality}. Among the variants, $\lambda_{\text{conf}} = 0.3$ consistently delivers strong predictive quality alongside efficiency. These results demonstrate that while \approach generalizes well across domains, its effectiveness hinges on thoughtful domain-specific fine-tuning and robust routing. 
\begin{table}[]
\renewcommand{\arraystretch}{1.3}
\caption{Comparison of \approach with Baseline LLMs for inter-domain setting (Set3)}
\resizebox{0.9\columnwidth}{!}{
\begin{tabular}{|lccc|}
\hline
\multicolumn{1}{|c|}{\textbf{\begin{tabular}[c]{@{}c@{}}Experimental\\ Candidates\end{tabular}}} & \multicolumn{1}{c|}{\textbf{\begin{tabular}[c]{@{}c@{}}Latency \\ (sec)\end{tabular}}} & \multicolumn{1}{c|}{\textbf{\begin{tabular}[c]{@{}c@{}}Confidence\\ Score (/5)\end{tabular}}} & \textbf{\begin{tabular}[c]{@{}c@{}}Energy\\ Consumption (kJ)\end{tabular}} \\ \hline
\multicolumn{1}{|l|}{Llama-2 13B}                                                                & \multicolumn{1}{c|}{1367.33}                                                           & \multicolumn{1}{c|}{\cellcolor[HTML]{F5B9B9}3.358}                                                           & \cellcolor[HTML]{F5B9B9}2.268                                              \\
\multicolumn{1}{|l|}{Deepseek-MOE 16B}                                                           & \multicolumn{1}{c|}{\cellcolor[HTML]{F5B9B9}1638.28}                                   & \multicolumn{1}{c|}{3.644}                                                                                   & 1.244                                                                      \\
\multicolumn{1}{|l|}{Qwen2.5-14B}                                                                & \multicolumn{1}{c|}{\cellcolor[HTML]{F5B9B9}1620.06}                                   & \multicolumn{1}{c|}{\cellcolor[HTML]{CFFFCA}3.946}                                                           & \cellcolor[HTML]{F5B9B9}2.591                                              \\ \hline
\multicolumn{4}{|c|}{\textbf{w/o Cache Module}}                                                                                                                                                                                                                                                                                                                                                \\ \hline
\multicolumn{1}{|l|}{\approach}                                                                  & \multicolumn{1}{c|}{\cellcolor[HTML]{CFFFCA}932.41}                                    & \multicolumn{1}{c|}{3.529}                                                                                   & 1.005                                              \\
\multicolumn{1}{|l|}{\approach ($\lambda_{lat}=0.3$)}                                            & \multicolumn{1}{c|}{\cellcolor[HTML]{CFFFCA}679.59}                                    & \multicolumn{1}{c|}{\cellcolor[HTML]{F5B9B9}3.214}                                                           & \cellcolor[HTML]{CFFFCA}0.855                                              \\
\multicolumn{1}{|l|}{\approach ($\lambda_{conf}=0.3$)}                                           & \multicolumn{1}{c|}{\cellcolor[HTML]{CFFFCA}969.07}                                    & \multicolumn{1}{c|}{\cellcolor[HTML]{CFFFCA}3.817}                                                           & 1.031                                              \\
\multicolumn{1}{|l|}{\approach ($\lambda_{energy}=0.2$)}                                           & \multicolumn{1}{c|}{\cellcolor[HTML]{CFFFCA}736.56}                                    & \multicolumn{1}{c|}{3.497}                                                           & \cellcolor[HTML]{CFFFCA}0.708                                              \\ \hline
\multicolumn{4}{|c|}{\textbf{with Cache Module (Cache Size = 3)}}                                                                                                                          \\ \hline
\multicolumn{1}{|l|}{\approach (CM)}                                                             & \multicolumn{1}{c|}{1454.87}                                                           & \multicolumn{1}{c|}{3.577}                                                                                   & 0.998                                              \\
\multicolumn{1}{|l|}{\approach (CM + $\lambda_{lat}=0.3$)}                                       & \multicolumn{1}{c|}{\cellcolor[HTML]{CFFFCA}946.84}                                    & \multicolumn{1}{c|}{\cellcolor[HTML]{F5B9B9}3.307}                                                           & \cellcolor[HTML]{CFFFCA}0.881                                              \\
\multicolumn{1}{|l|}{\approach (CM + $\lambda_{conf}=0.3$)}                                      & \multicolumn{1}{c|}{\cellcolor[HTML]{F5B9B9}1816.37}                                   & \multicolumn{1}{c|}{\cellcolor[HTML]{CFFFCA}3.802}                                                           & 1.057                                              \\
\multicolumn{1}{|l|}{\approach (CM + $\lambda_{energy}=0.2$)}                                       & \multicolumn{1}{c|}{\cellcolor[HTML]{CFFFCA}813.31}                                    & \multicolumn{1}{c|}{3.527}                                                           & \cellcolor[HTML]{CFFFCA}0.679                                          \\   
\hline
\end{tabular}
}
\label{tab:generalizability}
\vspace{-12pt}
\end{table}
\textbf{RQ3: To what extent does the system maintain efficiency
in terms of decision-making feasibility and runtime
overhead under resource constraints}

To evaluate system efficiency, we measure decision-making overhead and runtime performance including CPU memory usage \textcolor{black}{(referred as DRAM)}, inference latency and energy consumption. Our routing strategy, based on cosine similarity over MiniLM embeddings, requires loading the MiniLM model on the CPU, introducing a memory overhead of approximately \textbf{200MB}. This additional usage is modest when compared to the \textcolor{black}{VRAM} requirements of inference models, which range from \textbf{28--33~GB in  LLMs}. In contrast, our cache-aware deployment reduces \textcolor{black}{VRAM} usage to as low as 18.8~GB, amounting to approximately 30\% reduction. Thus, the CPU-side overhead remains minimal in the context of overall resource savings. Inference benefits from lightweight specialized models, reducing average per-query time to \textbf{5.2} seconds, nearly half of the \textbf{10.3} seconds observed with LLMs, thereby improving responsiveness under lower compute load. The energy consumption of our approach, measured across all components including model selection, caching, and inference, remains consistently lower than that of the LLM baselines. Despite the added decision-making step, the additional energy consumed is only \textbf{2J}, negligible compared to the 2–5 kJ used by LLMs. The step adds just \textbf{38} ms on average (see Table~\ref{tab:router}), confirming that routing and control introduce minimal latency, demonstrating that the added decision overhead doesn't negatively effect the efficiency of the approach. 

\color{black}
In our experiments, a fixed maximum cache size limits the number of resident models, independent of dynamic VRAM pressure. Each model load or eviction incurs a cold-start overhead of approximately 1.5–2 seconds, with frequency determined by the cache hit ratio. The self-adaptive caching mechanism explicitly accounts for this overhead and aims to minimize cold starts by maximizing cache hits as observed in Table ~\ref{tab:caching-adaptive}, higher $\lambda_{cache}$ corresponds to higher cache hit ratio implying fewer amount of cold starts.

For the LLM-as-a-Judge setup, we use a common prompt template with the temperature fixed to zero to ensure deterministic evaluation. To assess reliability, we conducted a human annotation study with two annotators. The LLM scores show moderate agreement with human judgments (Spearman $\rho$ = 0.484 and 0.541), while inter-annotator agreement is also moderate (quadratic Cohen’s $\kappa$ = 0.457). The average standard deviation of LLM confidence scores over three runs is 0.07, indicating stable evaluations.
\color{black}

\section{Discussion}
Our work, \approach, demonstrates that model organization and orchestration can matter more than model size. Coordinating lightweight, domain-specialized models via adaptive control reduced latency and resource usage without compromising performance. The combination of modular design and runtime feedback proved effective in balancing diverse goals, speed, performance, and efficiency while adapting to the inherent uncertainty of LM–based systems. These findings offer a foundation for practical deployment strategies and future research directions, discussed below.

\subsection{Implications for Practitioners}
\smallskip
\noindent \textit{6.1.1 Implications for LLM Consumers.} 
CALM directly mitigates \textbf{Challenge 1} where consumers require accurate, domain-specific outputs and have to rely on general purpose LLMs. These systems suffer from high latency, unpredictable output quality, and significant inference costs. \textbf{RQ1.1} shows that selecting smaller, domain-specialized models based on context can match the confidence levels of large LLMs while drastically reducing computational costs. \textbf{RQ1.3} further demonstrates that \approach dynamically balances speed and reliability in real time. For consumers in important domains like healthcare and law, this translates into faster, cheaper, and more reliable responses that are better aligned with user intent, improving trust and usability in domain-specific deployments.
\vspace{1pt}

\noindent \textit{6.1.2 Implications for LLM Providers.} 
As highlighted in \textbf{Challenge 2}, model providers have to bear the burden of high inference cost, latency and privacy concerns from external data transfer ~\cite{collins2024llmcost}. To address these, \textbf{RQ1.2} shows that integrating an LRU cache significantly reduces memory usage with minimal performance loss, making it feasible to run on constrained hardware like edge devices or shared GPUs. \textbf{RQ1.3} demonstrates that \approach's self-adaptive feedback loop allows real-time trade-offs between speed and reliability based on system load and task criticality. It adapts to two key sources of uncertainty: \textbf{(1)} inherent variability in LMs that affect latency and confidence, and \textbf{(2)} system-level uncertainty from dynamic SLM selection, which impacts memory and energy consumption (\textbf{Challenge 4}). Together, these mechanisms support SLA adherence across diverse workloads without overfitting to any single use case.
\textbf{RQ2} further confirms that the modular approach generalizes well across domains,  With the performance gap between SLMs and LLMs narrowing (studies show performance gap has reduced from $\sim$20\% to 2\%~\cite{gdowik2025slm_enterprise_ai}), domain-specialized models with \approach offer a scalable, cost-efficient solution to the deployment and management challenges of SLM fleets highlighted in \textbf{Challenge 3}.

\smallskip

\subsection{Implications for Research}

Prior studies~\cite{sens2025largescalestudymodelintegration, kwon2023pagedattention, sun2024llumnix} have addressed scalability and performance in LLM systems but focus on isolated techniques. \textbf{\approach} applies core SE principles ie, modularity, runtime adaptability, and resource-aware scheduling, to large-scale SLM deployment. By replacing single LLM setups with loosely coupled, domain-specialized SLMs, \textbf{RQ1.1} and \textbf{RQ1.3} show that modular design enables scalable, latency-sensitive deployment under dynamic workloads. LRU-based caching (\textbf{RQ1.2}) supports efficient model reuse under memory constraints, simulating an unbounded model pool. Unlike static approaches~\cite{modular2025serving}, \approach enables real-time tuning of latency-confidence trade-offs based on workload behavior and policy goals which can be extended to incorporate energy efficiency as a system-level objective instead of a passive outcome. Moreover, enabling user feedback on model responses introduces a human-in-the-loop mechanism that enables self-adaptation strategies to accomodate human preferences. These results highlight major insights into addressing \textbf{Challenges 2 and 3}, also offering two key directions for future research. First, in \textbf{SE for AI}, there is a need to develop control mechanisms that dynamically adapt model selection in response to changing goals, user intent, or resource availability, moving beyond fixed priorities to context-aware reconfigurable strategies. 
More work can be developed on uncertainty-aware routing strategies for LM-based systems, to automatically adjust the model preferences based on the overall load on the system.
Second, in \textbf{Sustainable AI} 
~\cite{fernandez2025energyconsiderationslargelanguage,jain2025dissectingtransformersclearperspective,JMLR:v24:23-0069} infrastructure, treating energy as a first-class system metric, designing routing policies that balance environmental and performance metrics, and a motivation to favor SLMs over LLMs in multi-agent, high-throughput, or workload-driven settings, could lead to significant cumulative reductions in energy impact.

\vspace{-6pt}
\section{Threats to Validity}

While we evaluate our approach across a wide range of ablations, some factors may affect internal validity. The routing mechanism relies on computing similarity between system descriptions and user-request, which makes it sensitive to minor variations in phrasing or structure. This could lead to inconsistent routing decisions under slightly altered inputs. Additionally, the control parameters used in our method are selected through empirical tuning and are influenced by the behavior of the base router, which may limit reproducibility. We note that under heavy load, increased queuing leads to elevated latency since the inference time per query is much higher than the rate of arrival of requests, which may obscure fine-grained latency differences across settings; running smaller-scale experiments could help isolate these effects more precisely.

External validity is shaped by the experimental assumptions and the simulation-based evaluation setup. We simulate user feedback using an LLM-as-a-Judge to assign confidence scores. While real user interaction could further validate our approach, the use of LLM-based feedback is an effective proxy, given the system's ability to adaptively optimize in real time from the provided scores. Furthermore, our evaluation is conducted on a fixed pool of SLMs fine-tuned for specific domains, offering a controlled setting to test the effectiveness of our method. We are optimistic that, with further evaluation across diverse model fleets and deployment settings, our routing and optimization strategy can generalize well and deliver robust performance in broader real-world applications. 

\vspace{-7pt}
\section{Related Works}
Over the years, extensive research has been done on self-adaptive systems~\cite{macias2013self,krupitzer2015survey,weyns2023self} and there have been different works that make use of ML techniques for performing self-adaptation~\cite{gheibi2021applying}. However, there have been some emerging works in the space of applying self-adaptation techniques to improve the QoS of AI-enabled systems. Casimiro et al.~\cite{casimiro2021self} highlighted challenges in ML-enabled systems and motivated self-adaptive solutions using MAPE-K feedback loop. They later proposed a probabilistic model-checking approach that adapts by reasoning about trade-offs between tactics when performing self-adaptation~\cite{casimiro2024self}. Sens et al.~\cite{kulkarni2023towards,marda2024switch} developed a self-adaptation approach that dynamically switching between different AI models to trade-off between different QoS parameters. However, the work does not take into consideration LLMs. Moreover, it considers all models to be available in memory to perform the switching. The same work was extended further to support energy efficiency by Tedla et al.~\cite{tedla2024ecomls} and Matathammal et al.~\cite{edgemlbalancer}. However, these works assume that all the models are in memory which has a significant impact on the overall efficiency and effectiveness. Moreover, they focus more on vision/ML based systems than on LLM-based systems. There has been ongoing work in managing fleets of domain-specific LMs instead of general purpose LLMs. BLADE~\cite{li2024bladeenhancingblackboxlarge} uses domain-specific SLM to achieve high performance on public legal and medical benchmarks but assumes a fixed LM and fails generalize to settings with multiple domains or limited deployment capacity. Routing frameworks like OrchestraLLM~\cite{lee-etal-2024-orchestrallm} improve efficiency by dispatching between LLMs and SLMs but lack self-adaptive routing based on high-level QoS goals and provide no scalable resource management for large SLM pools. TO-Router~\cite{stripelis-etal-2024-tensoropera} handles LLM expert routing but ignores SLM experts, which can deliver similar domain-specific performance at much lower resource costs. \approach overcomes these limitations by dynamically loading models, using SLMs for broader applicability, and enabling self-adaptive ML systems that adjust to changing QoS goals and resource constraints, going beyond fixed, domain assumptions and static routing through continuous decision re-evaluation.

\vspace{-7pt}
\section{Conclusion}
We presented \textbf{CALM}, a self-adaptive approach for managing domain-specialized SLMs with a focus on resource efficiency and QoS-aware decision-making. It addresses key limitations of prior approaches through dynamic model loading, adaptive routing, and efficient management of lightweight models across domains. Our evaluation shows that CALM improves performance, reduces latency and energy consumption compared to single-LLM setups with efficient deployment through lower memory footprint. The integrated feedback loop enables adaptation to shifting system goals, making CALM a practical solution for deploying AI systems under real-world constraints. While CALM demonstrates effective single-GPU deployment, future efforts could explore multi-GPU and distributed execution strategies to further scale SLM hosting and enhance throughput. Furthermore, improving the system’s resilience to minor variations in system descriptions and expanding its evaluation to multiturn conversational settings could enhance usability. Future research could include handling long or complex input prompts reliably, ensuring consistent performance across a wider range of real-world scenarios.

\bibliographystyle{ACM-Reference-Format}
\bibliography{references}

\end{document}